\documentclass[a4paper,12pt]{article}
\pdfoutput=1

\usepackage{
graphicx,
hyperref,
amsmath,
amssymb,
charter,
xcolor,
ifluatex,
cite,
longtable,
booktabs,
multirow,
bbold}

\usepackage{colortbl}

\definecolor{Gray}{gray}{0.95}
\definecolor{RGray}{gray}{0.85}
\definecolor{CGray}{gray}{0.92}

\newcommand{\Vv}{{\cal V}}
\newcommand{\Ss}{{\cal S}}
\newcommand{\Pp}{{\cal P}}
\newcommand{\Hh}{{\cal H}}
\newcommand{\C}{{\cal C}}
\newcommand{\D}{{\cal D}}
\newcommand{\U}{{\cal U}}
\newcommand{\E}{{\cal E}}
\newcommand{\N}{{\cal N}}
\newcommand{\M}{{\cal M}}

\newcommand{\B}{{\cal B}}
\renewcommand{\L}{{\cal L}}
\renewcommand{\O}{{\cal O}}

\newcommand{\eq}[1]{\begin{equation} #1 \end{equation}}
\newcommand{\eqa}[1]{\begin{eqnarray} #1 \end{eqnarray}}
\newcommand{\nvl}{n_{\scriptscriptstyle \rm VL}}
\newcommand{\vl}{vector-like }

\newcommand{\gev}{\text{ GeV}}
\newcommand{\ep}[1]{{\epsilon^#1}}
\newcommand{\hc}{\mathrm{h.c.}}

\newcommand{\Tab}[1]{Table~\ref{#1}}
\newcommand{\Sec}[1]{Section~\ref{#1}}
\newcommand{\Eq}[1]{Eq.~(\ref{#1})}
\newcommand{\pdg}{\cite{Agashe:2014kda}}
\newcommand{\hfag}{\cite{Amhis:2014hma}}

\newcommand{\skipp}[1]{}
\usepackage[normalem]{ulem}

\makeatletter

\hypersetup{colorlinks,bookmarksopen,bookmarksnumbered,linkcolor=blus,pdfstartview=FitH,urlcolor=blus,citecolor=verde}

\usepackage{multicol}
\definecolor{tit}{rgb}{0.1,0.2,0.4}
\definecolor{blus}{cmyk}{1,1,0,0.6}
\definecolor{verde}{cmyk}{0.92,0,0.59,0.25}

\begin{document}
\allowdisplaybreaks
\vspace*{-2.5cm}
\begin{flushright}
{\small
LMU-ASC 34/16\\
IFIC/16-62
}
\end{flushright}

\vspace{2cm}

\begin{center}
{\LARGE \bf \color{tit}
Phenomenology of an $\boldsymbol{\mathrm{SU(2)} \times \mathrm{SU(2)} \times \mathrm{U(1)}}$\\[3mm]
model with lepton-flavour non-universality}\\[1cm]

{\large\bf Sofiane M. Boucenna$^{a}$, Alejandro Celis$^{b}$, Javier Fuentes-Mart\'in$^{c}$,\\[2mm]
Avelino Vicente$^{c}$, Javier Virto$^{d}$}  
\\[7mm]
{\it $^a$ } {\em INFN, Laboratori Nazionali di Frascati, C.P. 13, 100044 Frascati, Italy}\\[3mm]
{\it $^b$ Ludwig-Maximilians-Universit\"at M\"unchen, 
   Fakult\"at f\"ur Physik,\\
   Arnold Sommerfeld Center for Theoretical Physics, 
   80333 M\"unchen, Germany}\\[3mm]
{\it $^c$ Instituto de F\'{\i}sica Corpuscular, Universitat de Val\`encia - CSIC, E-46071 Val\`encia, Spain}\\[3mm]
{\it $^d$ Albert Einstein Center for Fundamental Physics, Institute for Theoretical Physics,\\
University of Bern, CH-3012 Bern, Switzerland.}

\vspace{1cm}
{\large\bf\color{blus} Abstract}
\begin{quote}

We investigate a gauge extension of the Standard Model in light of the observed hints of lepton universality violation in $b \to c \ell \nu$ and $b \to s \ell^+ \ell^-$ decays at BaBar, Belle and LHCb.    The model consists of an extended gauge group $\mathrm{SU(2)}_{1} \times \mathrm{SU(2)}_{2} \times \mathrm{U(1)}_Y$ which breaks spontaneously around the TeV scale to the electroweak gauge group.  Fermion mixing effects with vector-like fermions give rise to potentially large new physics contributions in flavour transitions mediated by $W^{\prime}$ and $Z^{\prime}$ bosons. This model can ease tensions in $B$-physics data while satisfying stringent bounds from flavour physics, and electroweak precision data.  Possible ways to test the proposed new physics scenario with upcoming experimental measurements are discussed. Among other predictions, the ratios $R_M=\Gamma(B\to M\mu^+\mu^-)/\Gamma(B\to Me^+e^-)$, with $M = K^*, \phi$, are found to be reduced with respect to the Standard Model expectation $R_M \simeq 1$.

\end{quote}

\thispagestyle{empty}
\end{center}

\begin{quote}
{\large\noindent\color{blus} 
}

\end{quote}

\newpage

\tableofcontents

\setcounter{footnote}{0}

\section{Introduction}
\label{sec:intro}

The Standard Model (SM) of particle physics, based on the $\mathrm{SU(3)}_C \times \mathrm{SU(2)}_L \times \mathrm{U(1)}_Y$ gauge group, is an extremely successful theory that accounts for a wide range of high energy experiments at both the intensity and energy frontiers.  It is nevertheless a theory that is widely considered to be incomplete, and manifestations of new physics (NP) are expected to show up around the TeV scale.

A large class of particularly attractive NP theories consider extensions of the SM where its gauge group is embedded into a larger one which breaks to the SM (directly or via various steps) at or above the TeV scale. In this view, the SM is seen as an effective model valid at low energies. These constructions include Grand Unified Theories (GUT), composite models and string-inspired models. Interestingly, when the last breaking of the extended gauge group occurs around the TeV scale, a plethora of observables are generally predicted. In particular, flavour physics observables constitute a powerful probe to test these models due to the impressive precision and reach of current experiments.

In this article we present a detailed phenomenological analysis focused on flavour observables of a minimal extension of the SM electroweak gauge group to $\mathrm{SU(2)}_1\times \mathrm{SU(2)}_2\times \mathrm{U(1)}_Y$. We remain agnostic as to the origin of such a gauge group but assume it is broken around the TeV scale.
Models based on an extra $\mathrm{SU(2)}$ factor have been considered since a long time and constitute some of the most studied NP theories as they are predicted by various well-motivated frameworks, such as $\mathrm{SO(10)}$ or $\mathrm{E_6}$ GUTs. Depending on how the $\mathrm{SU(2)}$ and $\mathrm{U(1)}$ factors are identified, we can have for instance Left-Right~\cite{Mohapatra:1974hk} and Un-unified~\cite{Georgi:1989ic} schemes (for a general classification, cf. Ref.~\cite{Hsieh:2010zr}).
The extra $\mathrm{SU(2)}$ factor  implies the existence of new force carriers in the form of heavy partners of the SM $W$ and $Z$ bosons. In general, their couplings to matter are dictated by the choice of representations of the SM fields and the exotic new fields (if any). In any case, a rich phenomenology is predicted.

The model we will analyse was first presented in Ref.~\cite{Boucenna:2016wpr}. While the construction of the model has been motivated mainly by recent anomalies in $B$ decays, we will carry out here a generic analysis of the model and impose the constraints arising from these hints only as a secondary step.

The salient features of our model are summarised as follows:
\begin{itemize}
\item The extended gauge symmetry $\mathrm{SU(2)}_1 \times \mathrm{SU(2)}_2\times
\mathrm{U(1)}_Y$ spontaneously breaks at the TeV scale to the SM electroweak group following the pattern
\begin{equation*}
\mathrm{SU(2)}_1\times \mathrm{SU(2)}_2\times
\mathrm{U(1)}_Y\stackrel{\text{TeV}}{\longrightarrow}\mathrm{SU(2)}_L\times
\mathrm{U(1)}_Y\stackrel{\text{EW}}{\longrightarrow}
\mathrm{U(1)}_{\mbox{\footnotesize em}} \,.
\end{equation*}
\item The SM fields are all charged under one of the $\mathrm{SU(2)}$'s only, with the same quantum numbers they have in the SM, whereas newly introduced \vl fermions are charged similarly to the lepton and quark doublets but under the other $\mathrm{SU(2)}$ group.
\item Fermion mixing effects (facilitated by the same scalar field which breaks the original group) between the exotic and SM fermions act as a source of flavour non-universal vector currents by modulating the couplings of the SM fermions to the new gauge bosons.
\end{itemize}

Let us now briefly summarise the current $B$ anomalies.
Measurements of $b \to c \ell \nu$ transitions for different final state leptons can be used to test lepton flavour universality to a great precision given the cancellation of many sources of theoretical uncertainties occurring in ratios such as    
\begin{equation*}
R(D^{(*)}) = \frac{ \Gamma(B \rightarrow D^{(*)} \tau \nu )}{\Gamma(B \rightarrow D^{(*)} \ell \nu )}    \,,
\end{equation*}
with $\ell =e~\text{or}~\mu$.   The latest average of BaBar, Belle and LHCb measurements for these processes is $R(D) = 0.397 \pm 0.049$ and $R(D^*) = 0.316 \pm 0.019$, implying a  combined deviation from the SM at the $4\sigma$ level~\cite{Amhis:2014hma}.    Additionally, a measurement of the ratio  
\begin{equation*}
R_K = \frac{ \Gamma(B \rightarrow K \mu^+ \mu^-)}{\Gamma(B \rightarrow K e^+ e^-)}  \,,
\end{equation*}
performed by the LHCb collaboration in the low-$q^2$ region shows a $2.6\,\sigma$ deviation from the
SM, $R_K = 0.745^{+0.090}_{-0.074}\pm0.036$~\cite{Aaij:2014ora}.   This observable constitutes a clean probe of lepton non-universal new physics (NP) effects as many sources of uncertainty cancel in the ratio~\cite{Hiller:2003js,Guevara:2015pza,Bordone:2016gaq}.  Intriguingly,  departures from the SM have also been reported in $b \to s \mu^+ \mu^-$ decay observables such as branching fractions and angular distributions.  Global fits to $b \to s \ell^+ \ell^-$ data performed by different groups show a good overall agreement and obtain a consistent NP explanation of these departures from the SM with significances around the $4 \sigma$ level~\cite{Descotes-Genon:2013wba,Horgan:2013pva,Ghosh:2014awa,Hurth:2014vma,Altmannshofer:2014rta,Altmannshofer:2015sma,Descotes-Genon:2015uva,Hurth:2016fbr}.   While in the case of $b\to s \mu\mu$ observables the issue of hadronic uncertainties still raises some debate~\cite{Beaujean:2013soa,Descotes-Genon:2014uoa,Jager:2014rwa,Hurth:2013ssa,Lyon:2014hpa,Ciuchini:2015qxb}, it is clear that a common explanation to all anomalies is only possible in the presence of NP.

%%%%%%%%%%%%%%%%%%%%%%%%%%%%%%%%%%%%%%%%%%%%%%%%%%%%%%%%%%%%%%%%%%%%%%%%%
\begin{figure}
\centering
\includegraphics[width=13.0cm]{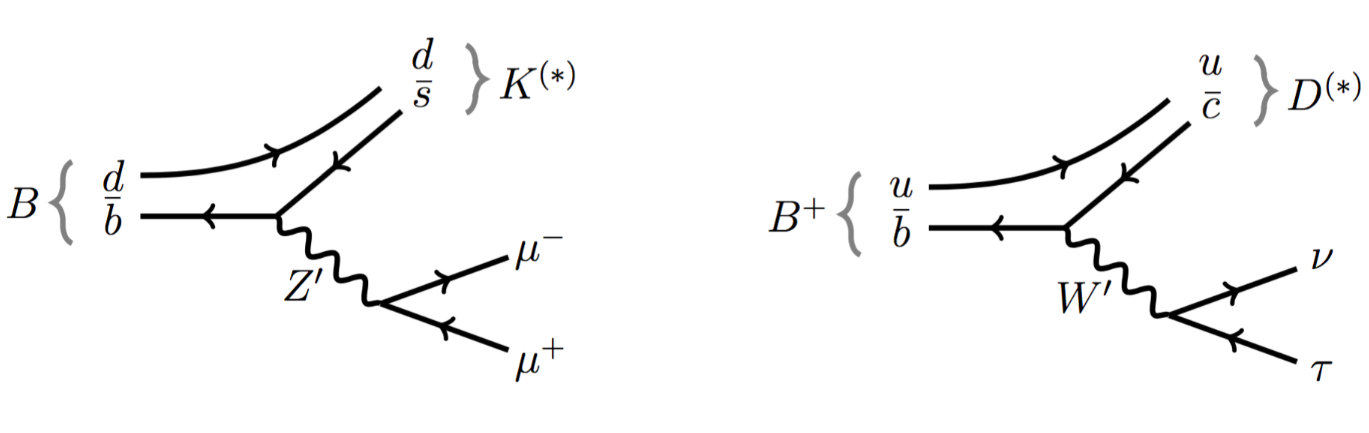}
\caption{\small \sf  New physics contributions to $B \to K^{(*)} \mu^+ \mu^-$ and $B \to D^{(*)} \tau \nu$ from the tree-level exchange of massive vector bosons.    }
\label{fig:diag1}
\end{figure}
%%%%%%%%%%%%%%%%%%%%%%%%%%%%%%%%%%%%%%%%%%%%%%%%%%%%%%%%%%%%%%%%%%%%%%%%%%

A considerable amount of efforts and model building activities have been devoted to these $B$-decay anomalies, 
though mainly focused on models that can accommodate only one of the
anomalies: either $R(D^{(*)})$ or $B \to K^{(*)} \ell^+ \ell^-$. The $R(D^{(*)})$ anomalies have been explained with charged scalars~\cite{Crivellin:2012ye,Celis:2012dk,Bailey:2012jg,Ko:2012sv,Crivellin:2015hha,Cline:2015lqp,Freytsis:2015qca,Nandi:2016wlp},
leptoquarks (or, equivalently, R-parity violating
supersymmetry)~\cite{Fajfer:2012jt,Deshpande:2012rr,Tanaka:2012nw,Sakaki:2013bfa,Dorsner:2013tla,Hati:2015awg,Zhu:2016xdg,Li:2016vvp}, or
a $W^{\prime}$ boson~\cite{He:2012zp}. Effects due to the presence of
light sterile neutrinos have also been explored in
Refs.~\cite{Abada:2013aba,Cvetic:2016fbv}. Models addressing the $B \to
K^{(*)} \ell^+ \ell^-$ anomalies on the other hand involved mostly a
$Z^{\prime}$ boson from an extended gauge
group~\cite{Buras:2013dea,Buras:2013qja,Altmannshofer:2014cfa,Crivellin:2015lwa,Crivellin:2015mga,Sierra:2015fma,Celis:2015ara,Belanger:2015nma,Celis:2015eqs,Falkowski:2015zwa,Allanach:2015gkd,Chiang:2016qov,Kim:2016bdu},
leptoquarks~\cite{Hiller:2014yaa,Biswas:2014gga,Gripaios:2014tna,Varzielas:2015iva,Becirevic:2015asa,Sahoo:2015wya,Sahoo:2015qha,Sahoo:2015fla,Pas:2015hca,Huang:2015vpt,Chen:2016dip},
or a massive resonance from a strong
dynamics~\cite{Niehoff:2015bfa,Niehoff:2015iaa,Carmona:2015ena}. In
contrast to these references, which rely on tree-level universality
violation, Ref.~\cite{Gripaios:2015gra} systematically explored
renormalizable models that explain $R_K$ at the 1-loop level.  The MSSM with R-parity was analysed in Ref.~\cite{Mahmoudi:2014mja}, finding that it is difficult to address the $b \to s \mu \mu$ anomalies. 

Unified explanations of both sets of anomalies are much more scarce.  This is due to the difficulty of accounting for
  deviations of similar size in processes that take place in the SM at
  different orders: loop level for $R_K$ and tree-level for
  $R(D^{(*)})$.   Nevertheless, among the proposed models we find those based on leptoquarks~\cite{Alonso:2015sja,Bauer:2015knc,Fajfer:2015ycq,Barbieri:2015yvd,Hati:2016thk,Deppisch:2016qqd,Das:2016vkr},
an extended perturbative gauge group~\cite{Boucenna:2016wpr},
or strongly-interacting models~\cite{Buttazzo:2016kid}. An effective field
theory approach has been adopted in
Refs.~\cite{Bhattacharya:2014wla,Greljo:2015mma,Calibbi:2015kma,Alonso:2015sja} and some observations
about the relevance of quantum effects have been given in
Ref.~\cite{Feruglio:2016gvd}.

In our model, the massive gauge vector bosons arising from the breaking of the extended gauge group mediate flavour transitions at tree-level as shown in Figure~\ref{fig:diag1}, providing a possible explanation to the deviations from the SM observed in $B$-meson decays~\cite{Boucenna:2016wpr}.

The plan of the paper is as follows: in \Sec{sec:model} we present the model in detail.    We derive the gauge boson and fermion masses and mixings, as well as the required textures in \Sec{sec:strategy}.     A detailed description of the flavour and electroweak observables included in the global fit is given in \Sec{sec:flavor}. Our global fit main results and predictions are presented in \Sec{sec:fit} and \Sec{sec:predictions}, respectively. Finally, in \Sec{sec:conclusions} we provide our conclusions.  Details of the model are provided in the Appendices.

\section{Description of the model}
\label{sec:model}

We consider a theory with the electroweak gauge group promoted to $\mathrm{SU(2)}_1 \times \mathrm{SU(2)}_2 \times \mathrm{U(1)}_Y$. 
The factor $\mathrm{U(1)}_Y$ corresponds to the usual hypercharge while the SM $\mathrm{SU(2)}_L$ is contained in the $\mathrm{SU(2)}$ product.   The gauge bosons and gauge couplings of the extended electroweak group will be denoted as:
\begin{align}
\begin{aligned}
\mathrm{SU(2)}_1 &: \quad g_1,\quad W^1_i\, , \\
\mathrm{SU(2)}_2 &: \quad g_2,\quad W^2_i \, , \\
\mathrm{U(1)}_Y &: \quad g^\prime,\quad B\,, \\
\end{aligned}
\end{align}
where $i=1,2,3$ is the $\mathrm{SU(2)}$ index.   All of the SM left-handed fermions transform exclusively under the second $\mathrm{SU(2)}$ factor, i.e.
\begin{align}
\begin{aligned}
q_L &= \left( {\bf 3} , {\bf 1} , {\bf 2} \right)_{\frac{1}{6}} \, ,  &   \ell_L &= \left( {\bf 1} , {\bf 1} , {\bf 2} \right)_{-\frac{1}{2}} \, , \\
u_R &= \left( {\bf 3} , {\bf 1} , {\bf 1} \right)_{\frac{2}{3}} \, ,  &   e_R &= \left( {\bf 1} , {\bf 1} , {\bf 1} \right)_{-1} \, , \\
d_R &= \left( {\bf 3} , {\bf 1} , {\bf 1} \right)_{-\frac{1}{3}} \, , &
\end{aligned}
\end{align}
where the representations refer to
$\mathrm{SU(3)}_C$, $\mathrm{SU(2)}_1$ and $\mathrm{SU(2)}_2$,  respectively, while the subscript denotes the hypercharge.
The SM doublets $q_L$ and $\ell_L$ can be
decomposed in $\mathrm{SU(2)}_2$ components in the usual way,
\begin{align}
\begin{aligned}
q_{L} & =
\begin{pmatrix}
u \\
d 
\end{pmatrix}_{L}
\, , \qquad
\ell_{L} =
\begin{pmatrix}
\nu \\
e 
\end{pmatrix}_{L} \, .
\end{aligned}
\end{align}
In addition, we introduce $\nvl$ generations of \vl
fermions transforming as
\begin{align}
\begin{aligned}
Q_{L,R} &\equiv\begin{pmatrix}
U \\
D 
\end{pmatrix}_{L,R} =\left( {\bf 3} , {\bf 2} , {\bf 1} \right)_{\frac{1}{6}} \, ;   
&   L_{L,R} &\equiv 
\begin{pmatrix}
N \\
E 
\end{pmatrix}_{L,R}= \left( {\bf 1} , {\bf 2} , {\bf 1} \right)_{-\frac{1}{2}} \, .
\end{aligned}
\end{align}
For the moment we take the number of generations $\nvl$ as a free parameter to be constrained by phenomenological requirements.
Symmetry breaking is achieved via the following set of scalars: a
self-dual bidoublet $\Phi$ (i.e., $\Phi =  \sigma^2 \Phi^\ast \sigma^2$, with $\sigma^2$ the usual Pauli matrix) and two doublets $\phi$ and $\phi'$,
\begin{align}
\begin{aligned}
\phi = \left( {\bf 1} , {\bf 1} , {\bf 2} \right)_{\frac{1}{2}} \,, \qquad 
\Phi = \left( {\bf 1} , {\bf 2} , {\bf \bar 2} \right)_{0} \, , \qquad
\phi^\prime = \left( {\bf 1} , {\bf 2} , {\bf 1} \right)_{\frac{1}{2}} \,,
\end{aligned}
\end{align}
which we decompose as:
\begin{align}
\phi=
\begin{pmatrix}
\varphi^+\\
\varphi^0
\end{pmatrix}\,,
\qquad\qquad
\Phi= \frac{1}{\sqrt{2}} \begin{pmatrix}
\Phi^0 & \Phi^+\\
-\Phi^- &  \bar \Phi^0
\end{pmatrix}\,,
\qquad\qquad
\phi^\prime=
\begin{pmatrix}
\varphi'^+\\
\varphi'^0
\end{pmatrix}\,,
\end{align}
with $\bar \Phi^0  = (\Phi^0)^*$ and $\Phi^- = \left( \Phi^+ \right)^\ast$. We summarise the particle content of the model in Table~\ref{tab:content}. 
\begin{table}
\centering
\begin{tabular}{| c c c c c c |}
\hline  
 & generations & $\mathrm{SU(3)}_C$ & $\mathrm{SU(2)}_1$ & $\mathrm{SU(2)}_2$ & $\mathrm{U(1)}_Y$ \\
\hline
\hline    
$\phi$ & 1 & ${\bf 1}$ & ${\bf 1}$ & ${\bf 2}$ & $1/2$ \\   \rowcolor{blue!15}
$\Phi$ & 1 & ${\bf 1}$ & ${\bf 2}$ & ${\bf \bar 2}$ & $0$ \\  \rowcolor{blue!15}
$\phi^\prime$ & 1 & ${\bf 1}$ & ${\bf 2}$ & ${\bf 1}$ & $1/2$ \\ 
\hline
\hline    
$q_L$ & 3 & ${\bf 3}$ & ${\bf 1}$ & ${\bf 2}$ & $1/6$ \\   
$u_R$ & 3 & ${\bf 3}$ & ${\bf 1}$ & ${\bf 1}$ & $2/3$ \\    
$d_R$ & 3 & ${\bf 3}$ & ${\bf 1}$ & ${\bf 1}$ & $-1/3$ \\     
$\ell_L$ & 3 & ${\bf 1}$ & ${\bf 1}$ & ${\bf 2}$ & $-1/2$ \\     
$e_R$ & 3 & ${\bf 1}$ & ${\bf 1}$ & ${\bf 1}$ & $-1$ \\  \rowcolor{blue!15}
$Q_{L,R}$ & $\nvl$ & ${\bf 3}$ & ${\bf 2}$ & ${\bf 1}$ & $1/6$ \\  \rowcolor{blue!15}
$L_{L,R}$ & $\nvl$ & ${\bf 1}$ & ${\bf 2}$ & ${\bf 1}$ & $-1/2$ \\
\hline
\hline
\end{tabular}
\caption{ \small \sf  Particle content of the model.  }
\label{tab:content}
\end{table}

\subsubsection*{Yukawa interactions}
\label{subsec:yukawa}

The SM fermions couple to the SM Higgs-like $\phi$ doublet with the usual Yukawa terms,
\begin{align}  \label{eqYl}
-\mathcal{L}_{\phi} &= \overline{q_L} \, y^{d} \, \phi \, d_R  + \overline{q_L} \, y^{u} \,  \tilde \phi \, u_R + \overline{\ell_L} \, y^{e} \, \phi \, e_R  + \hc \, ,
\end{align}
with $\tilde \phi \equiv i \sigma^2 \phi^*$. The $y^{u,d,e}$ Yukawa couplings represent $3\times3$ matrices in
family space. 
The  \vl fermions, on the other hand, have gauge-invariant Dirac mass terms,
\begin{align}   \label{eqmvl}
- \mathcal L_M =  \overline{Q_L}\, M_Q\, Q_R + \overline{L_L}\, M_L\, L_R + \hc \, ,
\end{align}
and our choice of representations  allows us to Yukawa-couple them to the SM fermions via
\begin{align}   \label{eqYh}
\begin{aligned}
- \mathcal L_{\Phi} &= \overline{Q_R}\, \lambda_q^\dagger\,\Phi\, q_L + \overline{L_R}\, \lambda_\ell^\dagger\, \Phi\, \ell_L + \hc \, ,
\end{aligned}
\end{align}
and
\begin{align}  \label{eqYl2}
- \mathcal{L}_{\phi^\prime} &= \overline{Q_L} \, \widetilde y^{d} \, \phi^\prime \, d_R  + \overline{Q_L} \, \widetilde y^{u} \,  \tilde \phi^{\prime}  \, u_R + \overline{L_L} \, \widetilde y^{e} \, \phi^\prime \, e_R  + \hc \, ,
\end{align}
where $\lambda_{q,\ell}$ and $\widetilde y^{u,d,e}$ are $3 \times \nvl$  and 
$\nvl\times3$ Yukawa matrices, respectively. After spontaneous symmetry breaking, these couplings will induce mixings between the \vl
and SM chiral fermions. This is crucial for the  phenomenology of the model,  in particular in its flavour sector, as will be clear in the next sections.

\subsubsection*{Scalar potential and symmetry breaking}
\label{subsec:scalar}

The scalar potential can be cast as follows:
\begin{align}
\begin{aligned}
\mathcal V &= m_\phi^2|\phi|^2 + \frac{\lambda_1}{2} |\phi|^4+m_{\phi^\prime}^2|\phi^{\prime}|^2 + \frac{\lambda_2}{2} |\phi^\prime|^4+ m_\Phi^2  \,\mathrm{Tr}(\Phi^{\dag} \Phi)   + \frac{\lambda_3}{2} \, \big[\mathrm{Tr}(\Phi^{\dag} \Phi)\big]^2\\
&\quad+\lambda_4 ( \phi^\dagger\phi ) (\phi^{\prime \dagger} \phi^\prime ) +  \lambda_5 (\phi^\dagger\phi )  \mathrm{Tr}(\Phi^{\dag} \Phi)  + \lambda_6 (\phi^{\prime \dagger}\phi^\prime )  \mathrm{Tr}(\Phi^{\dag} \Phi)+  \left(\mu \,  \phi^{\prime \dagger} \,  \Phi \, \phi  + \hc \right)\,.
\end{aligned}
\end{align}
We will assume that the parameters in the scalar potential are such
that the scalar fields develop vevs in the following directions:
\begin{align}
\langle \phi \rangle = \frac{1}{\sqrt{2}}
\begin{pmatrix} 
0 \\
v_\phi \end{pmatrix},
\quad \quad 
\langle \phi' \rangle = \frac{1}{\sqrt{2}}
\begin{pmatrix} 
0 \\
v_{\phi'} \end{pmatrix},
\quad \quad 
\langle \Phi \rangle = \frac{1}{2}
\begin{pmatrix} 
u & 0 \\
0 & u \end{pmatrix} \, .
\end{align}
Assuming $u\gg v_\phi , v_{\phi'}$, the symmetry breaking proceeds via the following pattern:
\begin{align}\label{eq:SSB_mI}
\mathrm{SU(2)}_1\times \mathrm{SU(2)}_2\times
\mathrm{U(1)}_Y\stackrel{u}{\longrightarrow}\mathrm{SU(2)}_L\times
\mathrm{U(1)}_Y\stackrel{v}{\longrightarrow}
\mathrm{U(1)}_{\mbox{\footnotesize em}}\,,
\end{align}
with the assumed vev hierarchy $u\sim\mbox{TeV}\gg v\simeq246$~GeV. With this breaking chain, the charge of the unbroken $\mathrm{U(1)}_{\rm{em}}$ group is defined as
\begin{equation}
Q = \left( T_3^1+T_3^2 \right) + Y = T_3^L + Y \, ,
\end{equation}
with $T_3^a$ the diagonal generator of $\mathrm{SU(2)}_a$.  In the first step, the original $\mathrm{SU(2)}_1\times \mathrm{SU(2)}_2$ group gets broken down to the diagonal $\mathrm{SU(2)}_L$. Under the diagonal sub-group, $\phi$ and $\phi'$ transform as doublets and, as usual with two-Higgs doublet models (2HDM), we parametrize their vevs as
\begin{align}
\begin{aligned}
v_\phi &= v \, \sin \beta \, , \\
v_{\phi'} &= v \, \cos \beta \, ,
\end{aligned}
\end{align}
where $v^2  = v_\phi^2 + v_{\phi'}^2$.
Since the two doublets transformed originally in a `mirror' way under the two original $\mathrm{SU(2)}$ factors, it is clear that the ratio between their vevs, $\tan \beta = v_\phi / v_{\phi'}$, controls the size of the gauge mixing effects. In particular, the limit $\tan \beta=g_1/g_2$ corresponds to the purely diagonal limit with no gauge mixing, see Subsection~\ref{ss:GB_mass_mix} for more details.

The scalar fields $\{\phi, \Phi,\phi^{\prime}\}$ contain 12 real
degrees of freedom, six of these become the longitudinal polarization
components of the $W^{(\prime) \pm}$ and $Z^{(\prime)}$ bosons.  In
the CP-conserving limit the scalar spectrum is composed of three
CP-even Higgs bosons, one CP-odd Higgs and one charged scalar, forming
an effective (constrained) 2HDM plus CP-even singlet system. The
scalar sector will present a decoupling behaviour, with a SM-like
Higgs boson at the weak scale (to be associated with the $125$~GeV
boson) and the rest of the scalars at the scale
$u\sim\mbox{TeV}$.\footnote{We will assume that $\mu$ is of the same
  order of the largest scale in the scalar potential, i.e. $\mu \sim
  u$.}  Further details of the scalar sector are given in
Appendix~\ref{sec:modeldetails}.

\section{Gauge boson and fermion masses and interactions}
\label{sec:strategy}
We now proceed to the analysis of the model presented in the previous section.   Here we will derive the masses and mixing of the gauge bosons and fermions of the model, as well as the neutral and charged vectorial currents.
  
\subsection{Fermion masses}

We can combine the SM and the \vl fermions as
\begin{eqnarray}
\U_{L,R}^I &\equiv& \left(u_{L,R}^i, U_{L,R}^k\right)  \,, \qquad
\D_{L,R}^I \ \equiv \ \left(d_{L,R}^i, D_{L,R}^k\right)\, , \nonumber \\[3mm]
\N_{L}^I &\equiv& \left(\nu_{L}^i, N_{L}^k\right) \, , \qquad\qquad
\N_{R}^I \ \equiv\ \left(0, N_{R}^k \right)\,,  \\[3mm]
\E_{L,R}^I &\equiv& \left(e_{L,R}^i, E_{L,R}^k\right)  \nonumber\, ,
\end{eqnarray}
where $i=1,2,3$, $k=1,\dots,\nvl$ and $I=1,\dots,3+\nvl$. With this
notation the fermion mass Lagrangian after symmetry breaking is given
by
\begin{equation}
-\L_{m}^f = \overline{\U_L} \M_\U \U_R + \overline{\D_L} \M_\D \D_R
+\overline{\E_L} \M_\E \E_R + \overline{\N_L} \M_\N \N_R + \hc
\end{equation}
The mass matrices are given in terms of the Yukawa couplings, \vl Dirac masses and vevs
as
\begin{equation}
\begin{array}{ll}
\M_\U = \left( \begin{array}{cc}
\frac{1}{\sqrt{2}} y_u v_\phi &  \frac{1}{2} \lambda_q u \\
\frac{1}{\sqrt{2}} \widetilde y_u v_{\phi^\prime} & M_Q
\end{array} \right) \,, \qquad 
&\M_\D = \left( \begin{array}{cc}
\frac{1}{\sqrt{2}} y_d v_\phi &  \frac{1}{2} \lambda_q u  \\
\frac{1}{\sqrt{2}} \widetilde y_d v_{\phi^\prime} & M_Q
\end{array} \right)\,, \\[8mm]
\M_\E = \left( \begin{array}{cc}
\frac{1}{\sqrt{2}} y_e v_\phi &  \frac{1}{2} \lambda_\ell u  \\
\frac{1}{\sqrt{2}} \widetilde y_e v_{\phi^\prime} & M_L
\end{array} \right)  \,, \quad 
&\M_\N = \left( \begin{array}{ccc}
0 &  \frac{1}{2} \lambda_\ell u  \\
0 & M_L
\end{array} \right) \, .
\end{array}
\end{equation}
Note that we did not include any mechanism to generate neutrino masses, and consequently $\M_\N$ leads to three massless neutrinos and $\nvl$ heavy  neutral Dirac fermions. It is nevertheless straightforward to account for neutrino masses without impacting our analysis and conclusions by including one of the usual mechanisms, such as the standard seesaw.

In order to have a manageable parameter space and simplify the analysis we will assume that the Yukawa couplings of $\phi^{\prime}$ can be neglected, $\widetilde y_{u,d,e}\simeq 0$.   This can be justified by introducing a softly-broken discrete $\mathcal{Z}_2$ symmetry under which $\phi^{\prime}$ is odd and all the other fields are even.   We take the Dirac masses of the vector-like fermions to be generically around the symmetry breaking scale $u\sim$ TeV.

The fermion mass matrices can be block-diagonalized perturbatively in the small ratio $\epsilon=v/u\ll1$ by means of the following field transformations
\begin{align}\label{eq:mass_basis}
\begin{aligned}
\U_L&\to V_Q^\dagger V_u^\dagger\,\U_L\,,\hspace{30pt} \U_R\,\to W_u^\dagger\, \U_R\,,\\
\D_L&\to V_Q^\dagger V_d^\dagger\,\D_L\,,\hspace{26pt} \D_R\,\to W_d^\dagger\, \D_R\,,\\
\E_L&\to V_L^\dagger V_e^\dagger\,\E_L\,,\hspace{32pt} \E_R\,\to W_e^\dagger\, \E_R\,,\\
\N_L&\to V_L^\dagger \,\N_L\,,
\end{aligned}
\end{align}
defined in terms of the unitary matrices
\begin{align}\label{eq:VQL}
V_{Q,L} &=
\left(
\begin{array}{c|c}
V_{Q,L}^{11} =  \sqrt{\mathbb{1}- \frac14 \lambda_{q,\ell} \widetilde M_{Q,L}^{-2}  \lambda_{q,\ell}^\dagger}
& V_{Q,L}^{12}=-\frac{u}{2} V_{Q,L}^{11} \lambda_{q,\ell} M_{Q,L}^{-1} \\[2mm]
\hline\\[-4mm]
V_{Q,L}^{21} = \frac{1}{2} \widetilde M_{Q,L}^{-1}  \lambda_{q,\ell}^\dagger &  V_{Q,L}^{22}=\frac{1}{u} \widetilde M_{Q,L}^{-1} M_{Q,L}^\dagger
\end{array}
\right)\,,\\[4mm]
V_f &= \mathbb{1} + i \epsilon^2 H_V^f + \dots
\quad ; \qquad
W_f \ =\  \mathbb{1} + i \epsilon H_W^f + \frac{(i \epsilon )^2}2 H_W^f{}^2 + \dots \label{eq:V1,W1} 
\end{align}
Here the freedom in the definition of $V_{Q,L}^{11}$ is removed by choosing it to be hermitian. Furthermore, $u\,\widetilde M_{Q,L}$ is the physical vector-like mass at leading order in $\epsilon$,
\begin{align}\label{eq:VL_mass}
\widetilde M_{Q,L} &=  \sqrt{\frac{M_{Q,L}^\dagger M_{Q,L}}{u^2} + \frac{\lambda_{q,\ell}^\dagger \lambda_{q,\ell}}{4}}
\simeq\mathrm{diag}\left(\widetilde M_{Q_1,L_1},\dots,\widetilde M_{Q_{\nvl},L_{\nvl}}\right)\,,
\end{align}
and the matrices $H_V^f$ and $H_W^f$ are given by
\begin{align}
H_V^f&= \frac{i}{2} \left(
\begin{array}{c|c}
0 & V_F^{11} y_f y_f^\dagger V_F^{21\dagger} \widetilde M_F^{-2} \\[2mm]
\hline\\[-4mm]
- \widetilde M_F^{-2} V_F^{21}y_f y_f^\dagger V_F^{11} &  0
\end{array}
\right)\,,\\[4mm]
H_W^f &= \frac{i}{\sqrt{2}} \left(
\begin{array}{c|c}
0 & y_f^\dagger V_F^{21\dagger} \widetilde M_F^{-1} \\[4mm]
\hline\\[-4mm]
- \widetilde M_F^{-1} V_F^{21}y_f &  0
\end{array}
\right)\,,
\end{align}
with $F=Q,L$ and $f=u,d,e$. After the block-diagonalization, a further diagonalization of the SM fermion block can be done by means of the $3\times3$ unitary transformations
\begin{align}\label{eq:3dmass_basis}
\begin{aligned}
u_L&\to S_u^\dagger u_L\,,\hspace{30pt} u_R\,\to U_u^\dagger\, u_R\,,\\
d_L&\to S_d^\dagger\,d_L\,,\hspace{29pt} d_R\,\to U_d^\dagger\, d_R\,,\\
e_L&\to S_e^\dagger\,e_L\,,\hspace{30pt} e_R\,\to U_e^\dagger\, e_R\,.
\end{aligned}
\end{align}
As in the SM, only one combination of these transformations appears in the gauge couplings: the CKM matrix, $V_{\rm CKM}=S_u S_d^\dagger$.

\subsection{Vector boson masses and gauge mixing}\label{ss:GB_mass_mix}

\subsubsection*{Neutral gauge bosons}

The neutral gauge bosons mass matrix in the basis $\Vv^0 = \left( W_3^1
, W_3^2 , B \right)$ is given by:
\begin{equation} 
\M_{\Vv^0}^2 = \frac{1}{4} \, \left( 
\begin{array}{ccc}
g_1^2 \left( v_{\phi^\prime}^2 + u^2 \right) & - g_1 g_2 u^2 & - g_1 g^\prime v_{\phi^\prime}^2 \\
- g_1 g_2 u^2 & g_2^2 \left( v_{\phi}^2 + u^2 \right) & - g_2 g^\prime v_{\phi}^2 \\
- g_1 g^\prime v_{\phi^\prime}^2 & - g_2 g^\prime v_{\phi}^2 & g'^2 \left( v_\phi^2 + v_{\phi^\prime}^2 \right)
\end{array} 
\right) \, .
\end{equation}
This matrix has one vanishing eigenvalue, corresponding to the photon and two massive eigenstates which are identified with the $Z$ and $Z^\prime$ bosons.  Before fully diagonalizing this mass matrix we consider first the rotation from $\left(W^1_3,\,W^2_3\right)$ to $\left(Z_h,\, W_3\right)$, with $W_3$ the electrically neutral $\mathrm{SU(2)}_L$ gauge boson. In order to do this we have to study the first symmetry breaking step, i.e. $u\neq0$ and $v=0$, diagonalize the top-left $2\times 2$ block and identify the massless state with $W_3$ (the $\mathrm{SU(2)}_L$ group remains unbroken in the first step). As a result we get:
\begin{align}
\begin{aligned}
Z_h&=\frac{1}{n_1}\left(g_1\, W^1_3 - g_2\, W^2_3\right)\,,\qquad \quad 
W_3=\frac{1}{n_1}\left(g_2\, W^1_3 + g_1\, W^2_3\right)\,,
\end{aligned}
\end{align}
with $n_1=\sqrt{g_1^2+g_2^2}$ and the gauge coupling of $\mathrm{SU(2)}_L$ taking the value  $g= g_1 g_2/n_1$. In the $\left(Z_h,\, W_3,\, B\right)$ basis, the rotation from $\left( W_3,\, B\right)$ to $\left(Z_l,\, A\right)$ is just like in the SM and we obtain:
\begin{align}
\begin{aligned}
Z_l&=\frac{1}{n_2}\left(g\, W_3 - g^\prime\, B\right)\,,\qquad \quad 
A=\frac{1}{n_2}\left(g^\prime\, W_3 + g\, B\right)\,,
\end{aligned}
\end{align}
where $n_2=\sqrt{g^2+g^{\prime\,2}}$ and the weak angle is defined as usual: $\hat{s}_W=g^\prime/n_2$ and $\hat{c}_W=g/n_2$. We are now in condition to write the neutral gauge boson mass matrix in the $\left(Z_h,\, Z_l,\, A\right)$ basis where it takes the form:
\begin{equation}
\mathcal M_{\Vv^0}^2 = \frac{1}{4} \,
\begin{pmatrix}
\left(g_1^2+g_2^2\right) u^2+\frac{g^2 g_2^2}{g_1^2}v^2 \left(   s_{\beta}^2  + \frac{g_1^4}{g_2^4}   c_{\beta}^2   \right)  & -g\,n_2\frac{g_2}{g_1}\,v^2  \left(    s_{\beta}^2   -   \frac{g_1^2}{g_2^2} c_{\beta}^2     \right)& 0\\[10pt]
-g\,n_2\frac{g_2}{g_1}\,v^2  \left(    s_{\beta}^2   -   \frac{g_1^2}{g_2^2} c_{\beta}^2     \right) & \left(g^2+g^{\prime \, 2}\right) v^2 & 0 \\
0 & 0 & 0
\end{pmatrix}\,.
\end{equation}
We see from this mass matrix that in the particular limit $v=0$, only $Z_h$ gets a mass $M_{Z'} =\tfrac{1}{4}\left(g_1^2+g_2^2\right) u^2$, which is expected since  $\mathrm{SU(2)}_L\times \mathrm{U(1)}_Y$ remains unbroken in that case. Moreover, we can extract the $Z_l-Z_h$ mixing.  The mass eigenvectors $\left(Z^\prime,\,Z\right)$ are given, in terms of $\left(Z_h,\,Z_l\right)$, by:
\begin{align}
\begin{aligned}
Z^\prime&=\cos\xi_Z\,Z_h-\sin\xi_Z\, Z_l\,,\qquad \quad 
Z=\sin\xi_Z\,Z_h+\cos\xi_Z\, Z_l\,,
\end{aligned}
\end{align}
with the mixing suppressed by the ratio $\epsilon\equiv v/u$,
\begin{align}
\xi_Z  \simeq\frac{g\,n_2}{n_1^2}\frac{g_2}{g_1}\,\epsilon^2  \left(  s_{\beta}^2   - \frac{g_1^2}{g_2^2}  c_{\beta}^2   \right)    =   \frac{g}{n_2}\frac{g_2}{g_1}\frac{M_Z^2}{M_{Z^\prime}^2}  \left(  s_{\beta}^2   - \frac{g_1^2}{g_2^2}  c_{\beta}^2   \right)   \,.
\end{align}
We define the parameter controlling the mixing as
\begin{equation}  \label{defmix}
\zeta =    s_{\beta}^2   - \frac{g_1^2}{g_2^2}  c_{\beta}^2    \,.
\end{equation} 
In the limit $\zeta \to 0$, the $\mathrm{SU(2)}_L$ sub-group corresponds to the diagonal subgroup of the original $\mathrm{SU(2)}$
product and gauge mixing vanishes. As anticipated in Section~\ref{subsec:scalar}, $\zeta\to 0$ corresponds to the limit $\tan\beta\to g_1/g_2$.

Finally, the masses of the neutral massive vector bosons are given by
\begin{align}
\begin{aligned}
M_{Z^\prime}^2&\simeq\frac{1}{4} \left(g_1^2+g_2^2\right) u^2    \,,\qquad \quad
M_Z^2\simeq\frac{1}{4}\left(g^2+g^{\prime\,2}\right)v^2\,.
\end{aligned}
\end{align}

\subsubsection*{Charged gauge bosons}

In the basis
$\Vv^+ = \left( W_{12}^1 , W_{12}^2 \right)$, with $W_{12}^r =
\frac{1}{\sqrt{2}} \left( W_1^r - i W_2^r \right)$, the charged gauge boson mass matrix is given by
\begin{equation} 
\M_{\Vv^+}^2 = \frac{1}{4} \, \left( 
\begin{array}{cc}
g_1^2 \left(v_{\phi^\prime}^2 + u^2\right) & -g_1 g_2 u^2 \\
-g_1 g_2 u^2 & g_2^2 \left(v_\phi^2 + u^2\right)
\end{array} 
\right) \, .
\end{equation}
As before, it is convenient to work in the basis
$\left(W_h,\,W_l\right)$ where the $\mathrm{SU(2)}_L$ gauge boson
appears explicitly.  To obtain this basis in terms of the
  original one, we set $v=0$, diagonalize the mass matrix and
  associate the null eigenvalue to $W_l$ ($\mathrm{SU(2)}_L$ remains
  unbroken in the first stage of symmetry breaking).   We get:
\begin{align}
\begin{aligned}
W_h&=\frac{1}{n_1}\left(g_1\, W^1 - g_2\, W^2\right)\,,\qquad \quad 
W_l=\frac{1}{n_1}\left(g_2\, W^1 + g_1\, W^2\right)\,.
\end{aligned}
\end{align}
In the basis $\left(W_h,\,W_l\right)$ the mass matrix reads:
\begin{align}
\M_{\Vv^+}^2 = \frac{1}{4} \, 
\begin{pmatrix}
\left(g_1^2 + g_2^2\right) u^2 +\frac{g^2 g_2^2}{g_1^2}v^2 \left(   s_{\beta}^2  + \frac{g_1^4}{g_2^4}   c_{\beta}^2   \right)  & -g^2 \frac{g_2}{g_1}v^2 \left(   s_{\beta}^2   - \frac{g_1^2}{g_2^2}  c_{\beta}^2      \right)   \\[5pt]
-g^2 \frac{g_2}{g_1}v^2 \left(   s_{\beta}^2   - \frac{g_1^2}{g_2^2}  c_{\beta}^2      \right)   & g^2v^2
\end{pmatrix} \, .
\end{align} 
The $W_l-W_h$ mixing presents the same structure as in the neutral gauge boson sector and reads:
\begin{align}
\xi_W \simeq \zeta\, \frac{g^2}{n_1^2}\frac{g_2}{g_1}\,\epsilon^2=  \zeta\, \frac{g_2}{g_1} \frac{M_W^2}{M_{W^\prime}^2}\,,
\end{align}
such that the physical eigenstates are given by:
\begin{align}
\begin{aligned}
W^\prime&=\cos\xi_W\,W_h-\sin\xi_W\, W_l\,,\qquad \quad 
W=\sin\xi_W\,W_h+\cos\xi_W\, W_l\,,
\end{aligned}
\end{align}
with masses
\begin{align}
\begin{aligned}
M_{W^\prime}^2  \simeq M_{Z^\prime}^2&\simeq  \frac{1}{4}  \left(g_1^2+g_2^2\right) u^2 \,,\qquad \quad 
M_W^2\simeq \frac{1}{4}g^2 v^2\,.
\end{aligned}
\end{align}

\subsection{Gauge boson couplings to fermions}

\subsubsection*{Neutral currents}
The neutral currents of the fermions are given by
\begin{align}
\begin{aligned}
\mathcal{L}_{\rm NC}&=\overline{\psi} \gamma_\mu \left( g' B^\mu Y + g_1 W_3^{1\mu} T_3^1 + g_2 W_3^{2\mu} T_3^2 \right) \psi\\
&=\overline{\psi} \gamma_\mu \left\{ e\, Q_\psi\,A^\mu + \frac{g}{c_W} Z_l^\mu [(T_3^1 + T_3^2) - s_W^2\, Q_\psi] + g Z_h^\mu \left[ \frac{g_1}{g_2} T_3^1 - \frac{g_2}{g_1} T_3^2 \right] \right\}\psi,
\end{aligned}
\end{align}
with $\psi=\U,\D,\E,\N$, and $e=g g^\prime/n_2$ and $Q_\psi$ denoting the electric coupling and the electric charge of the fermions, respectively. Applying the transformations in Eqs.~\eqref{eq:mass_basis} and~\eqref{eq:3dmass_basis} we can easily translate the above interactions to the fermion mass eigenbasis
\begin{align}
\begin{aligned}
\mathcal{L}_{\rm NC}\to\mathcal{L}_{\rm NC}&=A^\mu \,\overline{\psi}\gamma_\mu\,eQ_\psi\psi+ \frac{g}{c_W} Z_l^\mu\Big\{ \overline{\psi}\gamma_\mu\left[(T_3^1 + T_3^2)P_L - s_W^2 Q\right]\psi\\
&\quad-\frac{1}{2}\left(\overline{\D_R}\,\gamma_\mu\, O_R^{dd}\, \D_R-\overline{\U_R}\,\gamma_\mu\,O_R^{uu}\, \U_R+\overline{\E_R}\,\gamma_\mu\, O_R^{ee}\,\E_R-\overline{N_R}\,\gamma_\mu N_R\right)\Big\}  \\
&\quad+\,\frac{\hat{g}}{2} Z_h^\mu \Big[\overline{\D_L}\,\gamma_\mu\, O_L^Q\, \D_L-\overline{\U_L}\,\gamma_\mu\, V O_L^Q V^\dagger\, \U_L+\overline{\E_L}\,\gamma_\mu\, O_L^L\,\E_L-\overline{\N_L}\,\gamma_\mu\, O_L^L\, \N_L\\
&\quad-\,\frac{g_1^2}{g_2^2}\left(\overline{\D_R}\,\gamma_\mu\, O_R^{dd}\, \D_R-\overline{\U_R}\,\gamma_\mu\,O_R^{uu}\, \U_R+\overline{\E_R}\,\gamma_\mu\, O_R^{ee}\,\E_R-\overline{N_R}\,\gamma_\mu N_R\right)\Big]\\
&\quad+\mathcal{O}\left(\frac{m_f^2}{u^2}\right)\,.
\end{aligned}
\end{align}
Here $m_f$ denotes the mass of a SM fermion with $f=u,d,e$, and we introduced the following definitions: 
\begin{align}
\label{eq:OLR}
O_L^{Q,L}&\equiv
\begin{pmatrix}
\Delta^{q,\ell} & \Sigma\\
\Sigma^\dagger & \Omega^{Q,L}
\end{pmatrix}
=\mathbb{1}-\frac{g_1^2+g_2^2}{g_2^2}
\begin{pmatrix}
V_{Q,L}^{12}(V_{Q,L}^{12})^\dagger & V_{Q,L}^{12}(V_{Q,L}^{22})^\dagger\\[3mm]
V_{Q,L}^{22}(V_{Q,L}^{12})^\dagger & V_{Q,L}^{22}(V_{Q,L}^{22})^\dagger
\end{pmatrix}\,,\\[3mm]
O_R^{ff^\prime}&\equiv
\begin{pmatrix}
0 &  \hat \Sigma^f\\
\big(\hat \Sigma^{f^\prime}\big)^\dagger & 1
\end{pmatrix}
=
\begin{pmatrix}
0 & -\frac{m_f}{u}\big(V_F^{11}\big)^{-1}\big(V_F^{21}\big)^\dagger\widetilde{M}_F^{-1}\\
-\widetilde{M}_F^{-1}V_F^{21}\big(V_F^{11}\big)^{-1}\frac{m_{f^\prime}}{u} & 1
\end{pmatrix}\,,\\[3mm]
 V&=
\begin{pmatrix}
V_{\rm CKM} & 0\\
0 & 1
\end{pmatrix}
\,,
\end{align}
with $F = Q, L$, and finally $\hat{g}\equiv g g_2/g_1$. 

\subsubsection*{Charged currents}
Similarly, the charged currents take the following form
\begin{align}
\begin{aligned}
\mathcal{L}_{\rm CC}=&\,\frac{g_1}{\sqrt{2}}W^1_\mu\left[\overline{U}\gamma^\mu D +\overline{N }\gamma^\mu E \right]+\frac{g_2}{\sqrt{2}}W^2_\mu\left[\overline{u }\gamma^\mu\,P_L\,d +\overline{\nu }\gamma^\mu\,P_L\,e \right]+\hc\\
=&\,\frac{g}{\sqrt{2}}W_l^\mu\left[\overline{\U }\gamma_\mu P_L \D +\overline{\N }\gamma_\mu P_L\E +\overline{U }\gamma_\mu P_R\,D +\overline{N }\gamma_\mu P_R\,E \right]\\
&-\frac{g}{\sqrt{2}}W_h^\mu\left[\frac{g_2}{g_1}\left(\overline{u }\gamma_\mu P_L d +\overline{\nu }\gamma_\mu P_L e \right)-\frac{g_1}{g_2}\left(\overline{U }\gamma_\mu D +\overline{N }\gamma_\mu\,E \right)\right]+\hc\,,
\end{aligned}
\end{align}
and, in the fermion mass eigenbasis (see Eqs.~\eqref{eq:mass_basis} and~\eqref{eq:3dmass_basis}), we have
\begin{align}
\mathcal{L}_{\rm CC}\to\mathcal{L}_{\rm CC}=&\,\frac{g}{\sqrt{2}}W_l^\mu\left[\overline{\U_L}\,\gamma_\mu\,V \D_L +\overline{\N_L}\,\gamma_\mu\,\E_L +\overline{\U_R}\,\gamma_\mu\,O_R^{ud}\,\D_R+\overline{\N_R}\,\gamma_\mu\,O_R^{\nu e}\,\E_R\right] \nonumber \\
&-\frac{\hat{g}}{\sqrt{2}}W_h^\mu\left[\overline{\U_L}\,\gamma_\mu\, V O_L^Q\,\D_L +\overline{\N_L}\,\gamma_\mu\, O_L^L\, \E_L-\frac{g_1^2}{g_2^2}\left(\overline{\U_R}\,\gamma_\mu\,O_R^{ud}\,\D_R+\overline{\N_R}\,\gamma_\mu\,O_R^{\nu e}\,\E_R\right)\right] \nonumber\\
&+\hc+\mathcal{O}\left(\frac{m_f^2}{u^2}\right)\,.
\end{align}

\subsubsection*{Flavour textures for the gauge interactions}

In order to accommodate the hints of lepton universality violation from the recent anomalies in $B$ decays without being in tension with other bounds, we require negligible couplings of the new gauge bosons to the first family of SM-like leptons and a large universality violation among the other two.  We now derive the conditions on the number of generations of the exotic fermions to accommodate such constraints.

Using Eqs.~\eqref{eq:OLR} and~\eqref{eq:VQL}, the matrix $\Delta^{q,\ell}$, that parametrize NP contributions to the left-handed gauge interactions with SM fermions, can be readily written in the following form
\begin{align}\label{eq:lam_nVL1}
\Delta^{q,\ell}=\mathbb{1}-\frac{g_1^2+g_2^2}{4g_2^2}\lambda_{q,\ell}\widetilde{M}^{-2}\lambda_{q,\ell}^\dagger\,,
\end{align}
where the second term is the source of lepton non-universality induced by the mixings between the SM and vector-like fermions generated by the $\lambda_{q,\ell}$ Yukawa couplings. On the other hand, right-handed couplings involving SM fermions, controlled by $O_R^{ff^\prime}$, are mass suppressed and they can be neglected for the interactions we are considering.

If we consider the minimal scenario with $\nvl=1$, the Yukawa couplings $\lambda_{q,\ell}$ can be written generically as
\begin{align}
\lambda_{q,\ell}=\frac{2g_2}{n_1}\widetilde M_{Q,L}
\begin{pmatrix}
\Delta_{d,e}\\[1mm]
\Delta_{s,\mu}\\[1mm]
\Delta_{b,\tau}
\end{pmatrix}\,.
\end{align}
Here $\Delta_{d,e}$, $\Delta_{s,b}$ and $\Delta_{\mu, \tau}$ are free real parameters, and without loss of generality we have chosen an appropriate normalization factor to simplify the expression of $\Delta^{q,\ell}$. We have also ignored possible complex phases in the couplings since we are not interested in CP violating observables. From Eq.~\eqref{eq:lam_nVL1} it is then clear that, for $\nvl=1$, NP contributions to the left-handed gauge couplings to SM fermions are given by
\begin{align}  \label{eq:Delta_nVL1}
\Delta^{q,\ell}_{\nvl=1}&=
\begin{pmatrix}
1-(\Delta_{d,e})^2 & \Delta_{d,e}\Delta_{s,\mu} & \Delta_{d,e}\Delta_{b,\tau}\\[1mm]
\Delta_{d,e}\Delta_{s,\mu} & 1-(\Delta_{s,\mu})^2 & \Delta_{s,\mu}\Delta_{b,\tau}\\[1mm]
\Delta_{d,e}\Delta_{b,\tau} & \Delta_{s,\mu}\Delta_{b,\tau} & 1-(\Delta_{b,\tau})^2
\end{pmatrix}\,.
\end{align}
As we can see, in the limit of no gauge boson mixing, NP contributions to the first family of SM fermions can only be suppressed if we fix $\Delta_{d,e}\simeq1$ and $\Delta_{s,\mu},\,\Delta_{b,\tau}\ll1$ which then implies  approximate universal couplings for the second and the third families. Hence, we need at least two generations of vector-like fermions in order to have enough freedom to accommodate the observed hints of lepton universality violation.

In the rest of this article we will take the minimal setup consisting
of $\nvl=2$ since there is no compelling reason to assume
  additional \vl generations. Moreover, in order to reduce the number
of free parameters in the analysis we choose the following texture for
the Yukawa matrices $\lambda_{q,\ell}$:
\begin{align}\label{eq:lambda_text}
\lambda_{q,\ell}=\frac{2g_2}{n_1}
\begin{pmatrix}
\widetilde M_{Q_1,L_1} & 0\\
0 & \widetilde M_{Q_2,L_2}\,\Delta_{s,\mu}\\[2mm]
0 & \widetilde M_{Q_2,L_2}\,\Delta_{b,\tau}
\end{pmatrix}\,,
\end{align}
where, again, $\Delta_{s,b}$ and $\Delta_{\mu, \tau}$ are free real parameters and the normalization factor is chosen for convenience.\footnote{Note however that the free parameters have to satisfy the condition $(1 - g^2/g_2^2)(\Delta_{s,\mu}^2+\Delta_{b,\tau}^2) \leq1$ for consistency with 
Eq.~\eqref{eq:VL_mass}.} The left-handed currents, parametrized in terms of $O_L^{Q,L}$ (see Eq.~\eqref{eq:OLR}) now read
\begin{align}  \label{eq:lambdamy}
\Delta^{q,\ell}&=
\begin{pmatrix}
0 & 0 & 0\\
0 & 1-(\Delta_{s,\mu})^2 & \Delta_{s,\mu}\Delta_{b,\tau}\\[2mm]
0 & \Delta_{s,\mu}\Delta_{b,\tau} & 1-(\Delta_{b,\tau})^2
\end{pmatrix}
\,,\\[3mm]
\Sigma&=      \label{eq:lambdamy2}
\begin{pmatrix}
\frac{g_1}{g_2} & 0\\
0 & \Delta_{s,\mu}\sqrt{\frac{n_1^2}{g_2^2}-\Delta_{s,\mu}^2-\Delta_{b,\tau}^2}\\[2mm]
0 & \Delta_{b,\tau}\sqrt{\frac{n_1^2}{g_2^2}-\Delta_{s,\mu}^2-\Delta_{b,\tau}^2}
\end{pmatrix}
\,,\\[3mm]
\Omega^{Q,L}&=  \label{eq:lambdamy3}
\begin{pmatrix}
1- \frac{g_1^2}{g_2^2} & 0\\
0 & \Delta_{s,\mu}^2+\Delta_{b,\tau}^2-\frac{g_1^2}{g_2^2}\\
\end{pmatrix}
\,,
\end{align}
which, by construction, provide the desired patterns for the NP contributions to accommodate the data.\\

\section{Flavour constraints}
\label{sec:flavor}

We consider in our analysis flavour observables receiving new physics contributions at tree-level from the exchange of the massive vector bosons.     Additionally, we consider bounds from electroweak precision measurements at the $Z$ and $W$ pole which are affected in our model due to gauge mixing effects.         

Regarding electroweak precision observables at the $Z$ and $W$ pole, we use the fit to $Z$- and $W$-pole observables performed in Ref.~\cite{Efrati:2015eaa}.       The fit includes the observables listed in Tables~1 and 2 of~\cite{Efrati:2015eaa},
and provides mean values, standard deviations and the correlation matrix for the following parameters:
the correction to the $W$ mass ($\delta m$), anomalous $W$ and $Z$ couplings to leptons
($\delta g_L^{W\ell_i}$, $\delta g_{L,R}^{Z\ell_i}$) and anomalous $Z$ couplings to
quarks ($\delta g_{L,R}^{Z u_i}$, $\delta g_{L,R}^{Z d_i}$). The results for these
``pseudo-observables" can be found in Eqs.~(4.5-4.8) and Appendix~B of Ref.~\cite{Efrati:2015eaa}.   The relevant expressions for these pseudo-observables within our model are given in Appendix~\ref{sec:EWPD}.

We collect the list of flavour observables included in our analysis in \Tab{tableflavor} and describe them in more detail in the following subsections.

\begin{table}
\hspace{-34pt}
\renewcommand{\arraystretch}{1}
\setlength{\tabcolsep}{11pt}
\begin{tabular}{@{}ccccr@{}}
\toprule[1.6pt] 
\multicolumn{1}{l}{\bf Leptonic $\tau$ decays} & && \\ 
 \midrule
 Observable & Experiment & Correlation  & SM & Theory \\
 \midrule
 $\Gamma_{\tau \to e \nu\bar\nu}/\Gamma_{\mu \to e \nu\bar\nu} $ & $1.350(4)\cdot 10^{6}$ & \multirow{2}{1cm}{$0.45$} & $1.3456(5)\cdot 10^{6}$ &  Eq.~\ref{tau1} \\
 $\Gamma_{\tau \to \mu \nu\bar\nu}/\Gamma_{\mu \to e \nu\bar\nu}$ & $1.320(4)\cdot 10^{6}$ & & $1.3087(5)\cdot 10^{6}$ & Eq.~\ref{tau2} \\
\midrule[1.6pt] 
\multicolumn{1}{l}{\bf $\boldsymbol{d \to u}$ transitions} &&& \\ 
 \midrule 
 Observable & Experiment & Correlation & SM & Theory \\ 
  \midrule 
$\Gamma_{\pi \to \mu \nu}/\Gamma_{\pi \to e\nu}$ & $8.13(3)\cdot 10^3$ & \multirow{2}{1cm}{$0.49$} & $8.096(1)\cdot 10^{3}$ & Eq.~\ref{dtou1}  \\  
$\Gamma_{\tau \to \pi \nu}/\Gamma_{\pi \to e\nu}$& $7.90(5)\cdot 10^7$ &  & $7.91(1)\cdot 10^{7}$ & Eq.~\ref{dtou2} \\
\midrule[1.6pt] 
\multicolumn{1}{l}{\bf $\boldsymbol{s \to u}$ transitions} &&& & \\ 
 \midrule 
 Observable & Experiment & Correlation & SM & Theory \\ 
  \midrule 
 $\Gamma_{K \to \mu \nu}/\Gamma_{K \to e \nu}$ & $4.02(2)\cdot 10^4$ & \multirow{2}{*}{$\left[
\begin{array}{ccc}
\cdot & \cdot & \cdot \\
0.27 & \cdot & \cdot \\
0.01 & 0.00 & \cdot \\
\end{array}
\right]$} & $4.037(2)\cdot 10^{4}$ & Eq.~\ref{stou1} \\  
 $\Gamma_{\tau \to K \nu}/\Gamma_{K \to e \nu}$  & $1.89(3)\cdot 10^7$ &   & $1.939(4)\cdot 10^{7}$ & Eq.~\ref{stou2} \\
 $\Gamma_{K^+ \to \pi \mu \nu}/\Gamma_{K^+ \to \pi e \nu}$   & $0.660(3)$ &  & $0.663(2)$ & Eq.~\ref{stou3} \\
\midrule[1.6pt] 
\multicolumn{1}{l}{\bf $\boldsymbol{c \to s}$ transitions} &&& & \\ 
 \midrule 
 Observable & Experiment &   & SM & Theory \\ 
  \midrule  
$\Gamma_{D\to K \mu\nu}/\Gamma_{D \to K e\nu}$ & $0.95(5)$ & $(S=1.3)$ & $0.921(1)$ & Eq.~\ref{ctos1} \\ 
$\Gamma_{D_s\to\tau\nu}/\Gamma_{D_s\to\mu\nu}$ & $10.0(6)$ &  & $9.6(1)$ & Eq.~\ref{ctos2} \\ 
 \midrule[1.6pt]
 \multicolumn{1}{l}{\bf $\boldsymbol{b \to s}$ transitions} &&& & \\ 
 \midrule
 Observable & Experiment &   & SM  & Theory \\ 
\midrule
$\Delta M_s/\Delta M_d$ & $35.13(15)$ &   & $31.2(1.8)$ & Eq.~\ref{DeltaM} \\[1mm] 
\midrule
Coefficient & Fit \cite{Descotes-Genon:2015uva} & Correlation & SM & Theory \\ 
 \midrule 
 $\C_{9\mu}^\text{NP}$ &  $-1.1(0.2)$ & \multirow{4}{*}
 { $\left[
\begin{array}{cccc}
\cdot & \cdot & \cdot & \cdot \\
-0.08 & \cdot & \cdot & \cdot \\
0.10 & -0.10 & \cdot & \cdot \\
0.02 & 0.02 & 0.87 & \cdot \\
\end{array}
\right]$} & 0. & Eq.~\ref{eqwcsbtos} \\ 
$\C_{10\mu}^\text{NP}$ &  $+0.3(0.2)$ &   & 0. & Eq.~\ref{eqwcsbtos} \\ 
$\C_{9e}^\text{NP}$ & $-0.3(1.7)$ &  & 0. & Eq.~\ref{eqwcsbtos} \\ 
$\C_{10e}^\text{NP}$ & $+0.6(1.6)$ &  & 0. & Eq.~\ref{eqwcsbtos} \\[1mm] 
\midrule[1.6pt] 
\multicolumn{1}{l}{\bf $\boldsymbol{b \to c}$ transitions} &&& & \\ 
\midrule
Observable & Experiment & Correlation & SM & Theory \\ 
\midrule 
$\Gamma_{B \to D \mu \bar\nu}/\Gamma_{B \to D e \bar\nu}$ & $0.95(09)$ &
\multirow{2}{*}{$+0.51$} & $0.995(1)$ &  Eq.~\ref{btoc1}\\
$\Gamma_{B \to D^* \mu \bar\nu}/\Gamma_{B \to D^* e \bar\nu}$ & $0.97(08)$ &  & $0.996(1)$ & Eq.~\ref{btoc1}\\ 
\midrule
$R(D)$ & $0.397(49)$ & \multirow{2}{*}{$-0.21$} & $0.297(17)$ & Eq.~\ref{btoc2} \\ 
$R(D^*)$ & $0.316(19)$ &   & $0.252(3)$ & Eq.~\ref{btoc2} \\ 
\midrule
$\Gamma_{B\to X_c \tau \nu}/\Gamma_{B\to X_c e \nu}$ & $0.222(22)$ &  & $0.223(5)$ & Eq.~\ref{btoc3} \\ 
\bottomrule[1.6pt] 
\end{tabular}
\caption{\small \sf  List of flavour observables used in the fit.}
\label{tableflavor}
\end{table}

\subsection{Leptonic Tau decays}

Leptonic tau decays pose very stringent constraints on lepton flavour universality \cite{Pich:2013lsa}.
We consider the two decay rates $\Gamma(\tau \to \{e,\mu\} \nu \bar \nu)$, normalized to the muon
decay rate to cancel the dependence on $G_F$.
We take the individual experimental
branching ratios and lifetimes from the PDG~\pdg. For the branching ratios we take the result of
the constrained fit, which gives a correlation of $14\%$ between both measurements. Once normalized to the
$\tau$ lifetime, the decay rates have a correlation of $45\%$, while the normalization to the muon decay rate
has a minor impact on the correlation of the ratios because its uncertainty is negligible. The experimental
results are summarized in \Tab{tableflavor}.

In our model, we have:
\eqa{
\label{tau1}
\frac{\Gamma(\tau \rightarrow e  \nu  \bar \nu )}{ \Gamma(\mu \rightarrow  e   \nu  \bar \nu )  }   &=&
\frac{\sum_{i,j} |\C^{e\tau}_{ij}|^2}{\sum_{i,j} |\C^{e\mu}_{ij}|^2} \times
\frac{m_{\tau}^5\,f( x_{e\tau})}{m_{\mu}^5\,f( x_{e\mu})}
\ ,\\[3mm]
\label{tau2}
\frac{\Gamma(\tau \rightarrow  \mu  \nu  \bar \nu )}{ \Gamma(\mu \rightarrow  e   \nu  \bar \nu )  }  &=&
\frac{\sum_{i,j} |\C^{\mu\tau}_{ij}|^2}{\sum_{i,j} |\C^{e\mu}_{ij}|^2} \times
\frac{m_{\tau}^5\,f( x_{\mu\tau})}{m_{\mu}^5\,f( x_{e\mu})}
\ ,  
}
where $x_{\ell \ell'} = m_\ell^2/m_{\ell'}^2$ and $f(x) = 1- 8 x + 8 x^3 - x^4 -12 x^2 \ln x $.
The Wilson coefficients $\C^{\ell_a \ell_b}_{ij}$ are given by
\eqa{
\label{cijll}
\C^{\ell_a \ell_b}_{ij} &=& \frac{4 G_F}{\sqrt{2}} \delta_{aj} \delta_{ib}
+\frac{\hat g^2}{4 M_{W'}^2} \bigg[ 2 \Delta^\ell_{aj} \Delta^\ell_{ib} - \Delta^\ell_{ab} \Delta^\ell_{ij}
+ \zeta (\Delta^\ell_{ab} \delta_{ij} + 2 \Delta^\ell_{aj} \delta_{ib} + 2 \delta_{aj} \Delta^\ell_{ib}) \bigg] \,. }
The resulting predictions in the SM can be found in \Tab{tableflavor}.  Leading radiative corrections and $W$-boson propagator effects are included in the SM predictions~\cite{Kinoshita:1958ru,Berman:1958ti,Marciano:1988vm,Marciano:1999ih}.

\subsection{$d\to u$ transitions}

We consider the decay rates $\Gamma(\pi\to\mu\nu)$ and $\Gamma(\tau\to\pi\nu)$,
normalized to $\Gamma(\pi\to e\nu)$ in order to cancel the dependence on the
combination $G_F |V_{ud}| f_\pi$. These ratios constitute important constraints on
flavour non-universality in $d\to u \ell \nu$ transitions.

We calculate the experimental values for these ratios taking the averages for
branching fractions and lifetimes from the PDG~\pdg, and imposing the constraint
$\B(\pi \to e \nu)+\B(\pi \to \mu \nu) =1$. We find a correlation of $49\%$ between both ratios.
The corresponding results are summarized in \Tab{tableflavor}.

The model predictions for these ratios are:
\eqa{
\label{dtou1}
\dfrac{\Gamma(\pi \rightarrow \mu \bar \nu) }{ \Gamma(\pi \rightarrow e \bar \nu) }  &=& \dfrac{\sum_j |\C_{2j}^{ud}|^2}{\sum_j |\C_{1j}^{ud}|^2}  \times
\left[\dfrac{\Gamma(\pi \rightarrow \mu  \bar \nu) }{ \Gamma(\pi \rightarrow e \bar \nu) }  \right]_{\mbox{\scriptsize{SM}}} \ , \\[2mm]
\label{dtou2}
\dfrac{\Gamma(\tau \rightarrow \pi \nu) }{ \Gamma(\pi \rightarrow e \bar \nu) }    &=& \dfrac{\sum_j |\C_{3j}^{ud}|^2}{\sum_j |\C_{1j}^{ud}|^2}\times
\left[\dfrac{\Gamma(\tau \rightarrow \pi \nu) }{ \Gamma(\pi \rightarrow e \bar \nu) }  \right]_{\mbox{\scriptsize{SM}}} \ , 
}
where the Wilson coefficients $\C_{ab}^{u_id_j}$ are given by
\eqa{
\label{cijud}
\C_{ab}^{u_id_j} &=& \frac{4 G_F}{\sqrt{2}} V_{ij} \delta_{ab}
+\frac{\hat g^2}{2 M_{W'}^2} \bigg[  (V \Delta^q)_{ij} \Delta_{ab}^\ell
- \zeta \big(V_{ij} \Delta^\ell_{ab}  + (V \Delta^q)_{ij} \delta_{ab}\big) \bigg]  \,.
}
For the SM contributions we follow Ref.~\cite{Pich:2013lsa}. We have:
\eqa{
\label{pienuSM}
\left[\frac{ \Gamma(  \pi \to e \bar \nu )  }{  \Gamma(   \pi \to \mu \bar \nu )}\right]_\text{SM}
&=& \frac{  m_e^2 }{   m_{\mu}^2  }      \left[   \frac{   1- m_{e}^2/m_{\pi}^2    }{     1- m_{\mu}^2/m_{\pi}^2   } \right]^2 (1+ \delta R_{\pi \to e/\mu})\ , \\[2mm]
\label{taupinuSM}
\left[\frac{  \Gamma(\tau \to \pi \nu) }{\Gamma(\pi \to \mu \nu)}\right]_\text{SM}  &=&
\frac{  m_{\tau}^3 }{ 2 m_{\pi}   m_{\mu}^2 } \left[   \frac{ 1- m_{\pi}^2/m_{\tau}^2 }{  1- m_{\mu}^2/m_{P}^2   } \right]^2 (1+ \delta R_{\tau/\pi})\ .
}
The calculation of $\delta R_{\pi \to e/\mu}$ relies on Chiral Perturbation Theory to order $\O(e^2 p^2)$ \cite{Cirigliano:2007ga}. The radiative correction factor $\delta R_{\tau/\pi}$
can be found in Ref.~\cite{Decker:1994ea}. The SM predictions for both ratios are collected
in \Tab{tableflavor}.

\subsection{$s\to u$ transitions}

We consider the decay rates $\Gamma(K\to\mu\nu)$ and $\Gamma(\tau\to K \nu)$,
normalized to $\Gamma(K\to e\nu)$ in order to cancel the dependence on the
combination $G_F |V_{us}| f_K$, as well as the semileptonic ($K_{\ell 3}$) ratio $\Gamma(K^+\to\pi^0\mu^+\nu)/\Gamma(K^+\to \pi^0 e^+ \nu)$.
These ratios pose also important constraints on flavour non-universality.

We take the experimental values for the decay rates $\Gamma(K^+\to\mu^+\nu)$, $\Gamma(K^+\to\pi^0 e^+\nu)$ and $\Gamma(K^+\to \pi^0 \mu^+ \nu)$ from the constrained fit to $K^+$ decay data done by the PDG~\pdg, including the correlation matrix. The correlation between $\Gamma(K^+\to\mu^+\nu)$ and $\Gamma(K^+\to e^+\nu)$ is calculated comparing the averages for the individual rates with the ratio given by the PDG, resulting in a correlation of $60\%$. Assuming no correlation between
$\Gamma(K^+\to e^+\nu)$ and the semileptonic modes, and assuming that the
$\tau$ mode is uncorrelated to the $K$ modes, we construct a $5\times 5$ correlation
matrix and calculate the three ratios of interest, including their $3\times 3$
correlation matrix. These results are collected in \Tab{tableflavor}.

The model predictions for these ratios are:
\eqa{
\label{stou1}
\dfrac{\Gamma(K \rightarrow \mu \bar \nu) }{ \Gamma(K \rightarrow e \bar \nu) }  &=& \dfrac{\sum_j |\C_{2j}^{us}|^2}{\sum_j |\C_{1j}^{us}|^2} \times
\left[\dfrac{\Gamma(K \rightarrow \mu  \bar \nu) }{ \Gamma(K \rightarrow e \bar \nu) }  \right]_{\mbox{\scriptsize{SM}}} \ , \\[2mm]
\label{stou2}
\dfrac{\Gamma(\tau \rightarrow K \nu) }{ \Gamma(K \rightarrow e \bar \nu) }  &=& \dfrac{\sum_j |\C_{3j}^{us}|^2}{\sum_j |\C_{1j}^{us}|^2}  \times  
\left[\dfrac{\Gamma(\tau \rightarrow K \nu) }{ \Gamma(K \rightarrow e \bar \nu) }  \right]_{\mbox{\scriptsize{SM}}} \ ,  \\[2mm]
\label{stou3}
\dfrac{\Gamma(K^+ \rightarrow  \pi  \mu \bar \nu) }{ \Gamma(K^+ \rightarrow  \pi   e  \bar \nu) }  &=& \dfrac{\sum_j |\C_{2j}^{us}|^2}{\sum_j |\C_{1j}^{us}|^2} \times
\left[\dfrac{\Gamma(K^+ \rightarrow  \pi  \mu \bar \nu) }{ \Gamma(K^+ \rightarrow  \pi   e  \bar \nu) }   \right]_{\mbox{\scriptsize{SM}}}  \,,
}
with the Wilson coefficients $\C_{ij}^{us}$ given in \Eq{cijud}.
The SM contributions for the first two ratios are given by the analogous expressions
to Eqs.~(\ref{pienuSM},\ref{taupinuSM})~\cite{Cirigliano:2007ga,Decker:1994ea}. The SM contributions
to $K_{\ell 3}$ are given by \cite{Cirigliano:2008wn,Cirigliano:2011ny}
 \eq{
 \dfrac{\Gamma(K^+ \rightarrow  \pi  \mu \bar \nu) }{ \Gamma(K^+ \rightarrow  \pi   e  \bar \nu) }   = \frac{  I^{(0)}_{K \mu}(\lambda_i)   \left(1 +    \delta_{\mbox{\scriptsize{EM}}}^{K \mu}   + \delta_{\mathrm{SU(2)}}^{K \pi}  \right)   }{I^{(0)}_{K e}(\lambda_i)   \left(1 +    \delta_{\mbox{\scriptsize{EM}}}^{K e}   + \delta_{\mathrm{SU(2)}}^{K \pi}  \right) }\ ,
}
where quantities $I^{(0)}_{K\ell}(\lambda_i)$, $\delta_\text{EM}^{K\ell}$,
$\delta_{\text{SU}(2)}^{K\pi}$ encoding phase-space factors, electromagnetic and isospin corrections can be found in Refs.~\cite{Kastner:2008ch,Cirigliano:2008wn,Cirigliano:2011ny}.
The numerical results for the SM contributions are collected in \Tab{tableflavor}.

\subsection{$c\to s$ transitions}

We consider the ratios 
$\Gamma(D\to K \mu\nu)/\Gamma(D\to K e\nu)$
and
$\Gamma(D_s\to\tau\nu)/\Gamma(D_s\to\mu\nu)$,
constraining respectively $\mu - e$ and $\tau - \mu$ non-universality.

For $D\to K \ell \nu$, we consider charged and neutral modes separately.
For $D^+\to \bar K^0 \ell^+ \nu$ we take the separate branching ratios from the PDG
assuming no correlation. For  $D^0\to K^- \ell^+ \nu$ we take the results from the
PDG constrained fit, including the $5\%$ correlation. We construct the $D^+$ and $D^0$
ratios separately, obtaining
$\Gamma(D^+\to \bar K^0 \mu^+\nu)/\Gamma(D^+\to \bar K^0 e^+\nu) = 1.05(9)$ and
$\Gamma(D^0\to K^- \mu^+\nu)/\Gamma(D^0\to K^- e^+\nu) = 0.93(4)$.
These two ratios, corresponding to the same theoretical quantity
(isospin-breaking effects are neglected here), are combined according to the PDG
averaging prescription. Since there is a $\sim1\,\sigma$ tension between both results, we
rescale the error by the factor $S=1.3$.

For $D_s\to \ell \nu$ we take the individual branching fractions from the PDG,
assuming no correlation. The resulting experimental numbers for both ratios are
collected in \Tab{tableflavor}.

The model predictions for these ratios are:
\eqa{
\label{ctos1}
\frac{\Gamma(D \to  K \mu \bar \nu)}{\Gamma(D \to  K e \bar \nu)} &=&
\dfrac{\sum_j |\C_{2j}^{cs}|^2}{\sum_j |\C_{1j}^{cs}|^2}  \times 
\left[\frac{ \Gamma(D \to  K \mu \bar \nu)}{\Gamma(D \to  K e \bar \nu)}   \right]_{\mbox{\scriptsize{SM}}} \ , \\[2mm]
\label{ctos2}
\frac{ \Gamma(  D_s \to \tau \bar \nu ) }{\Gamma(  D_s \to \mu \bar \nu )  } &=&
\dfrac{\sum_j |\C_{3j}^{cs}|^2}{\sum_j |\C_{2j}^{cs}|^2} \times 
\left[\frac{ \Gamma(D_s \to \tau \bar \nu)}{\Gamma(D_s \to \mu \bar \nu)}  \right]_{\mbox{\scriptsize{SM}}} \ ,
}
with the Wilson coefficients $\C_{ij}^{cs}$ given in \Eq{cijud}.

Our SM prediction for the leptonic decay modes includes electromagnetic corrections following~\cite{Fajfer:2015ixa}.    For the SM prediction of the semileptonic modes we use the BESIII determination of the form factor parameters in the simple
pole scheme as quoted in HFAG~\hfag.       The resulting SM predictions
are given in \Tab{tableflavor}.

\subsection{$b\to s$ transitions}

We consider here $b\to s$ transitions that are loop-mediated in the SM
but receive NP contributions at tree-level in our model (via $Z'$ and
$Z$ with anomalous couplings).  To the level of precision we are
working, the normalization factors in the SM amplitude ($G_F$ and CKM
elements) can be taken from tree-level determinations within the SM,
and it is not necessary in this case to consider only ratios where
these cancel out.

\subsubsection*{Mass difference in the $B_s$ system}

The observable $\Delta M_s$ constitutes a strong constraint on the $Z' s b$ coupling, independent of
the coupling to leptons. In order to minimize the uncertainty from hadronic matrix elements, we consider
the ratio $\Delta M_s/\Delta M_d$. We note that within our model set-up, $\Delta M_d$ does not receive
NP contributions at tree-level.

The experimental value for the ratio is obtained from the individual measurements for $\Delta M_{d,s}$,
which are known to subpercent precision~\hfag.
The result is given in \Tab{tableflavor}.

The theory prediction is given by:
\eq{
\label{DeltaM}
\frac{\Delta M_s}{\Delta M_d} =
\frac{M_{B_s}}{M_{B_d}} \,\xi^2\,\left|\frac{\C_{sb}}{\C_{db}} \right| =
\frac{M_{B_s}}{M_{B_d}} \,\xi^2\,\left| \frac{V_{ts}^2}{V_{td}^2} + \frac{\C_{sb}^\text{NP}}{\C_{db}^\text{SM}} \right|\ ,
}
where the Wilson coefficients $\C_{d_ib}=\C_{d_ib}^\text{SM}+\C_{d_ib}^\text{NP}$ are given by
%in \Eq{cib}, and
%
\eqa{
\label{cib}
 \C_{d_ib}^\text{SM}   &=& \frac{G_F^2 M_W^2}{4\pi^2} (V_{ti}V^\star_{tb})^2 S_0(x_t) \,, \qquad \C_{d_ib}^\text{NP} =   \frac{\hat g^2}{8 M_{W'}^2} (\Delta^q_{i3})^2  \,.
}
Here $S_0(x_t) = 2.322 \pm 0.018$ is the loop function in the SM~\cite{Lenz:2010gu}.    The parameter $\xi^2 = f_{B_s}^2B_{B_s}^{(1)}/f_{B_d}^2B_{B_d}^{(1)}$ is a ratio of decay constants and matrix elements
determined from lattice QCD. We consider the latest determination of the parameter $\xi$ from the FNAL/MILC collaborations~\cite{Bazavov:2016nty}:
$\xi = 1.206(18)(6)$.   The SM prediction is given by the first term in \Eq{DeltaM} and results in $(\Delta M_s/\Delta M_d)_\text{SM} = 31.2(1.8)$.

\subsubsection*{$b\to s\ell\ell$ observables}

We consider all $b\to s \ell \ell$ observables used in the fit of Ref.~\cite{Descotes-Genon:2015uva}:
\begin{itemize}
\item Branching ratios for $B\to X_s \mu^+\mu^-$ and $B_s\to \mu^+\mu^-$~\cite{Aubert:2004it,Iwasaki:2005sy,Huber:2015sra,CMS:2014xfa,Bobeth:2013uxa}.
\item Branching ratios for $B\to K e^+e^-$ (in the bin $[1,6]\gev^2$) and $B\to K \mu^+\mu^-$ (both at low and high $q^2$)~\cite{Aaij:2014tfa,Aaij:2014ora}.
\item Branching ratios, longitudinal polarization fractions and optimized angular observables~\cite{Matias:2012xw,DescotesGenon:2012zf,Descotes-Genon:2013vna} for
$B\to K^* e^+e^-$ (at very low $q^2$) and $B\to K^* \mu^+\mu^-$, $B_s\to \phi \mu^+\mu^-$ (both at low and high $q^2$)~\cite{Aaij:2013iag,Aaij:2013hha,Aaij:2014pli,Aaij:2015oid,Aaij:2013aln,Aaij:2015esa,Aaij:2015dea}.
\end{itemize}
Definitions, theoretical expressions and discussions on theoretical uncertainties can
be found in Refs.~\cite{Descotes-Genon:2013vna,Descotes-Genon:2015uva}. We follow the approach of
Ref.~\cite{Descotes-Genon:2014uoa} for $B\to V$ form factors, and take into account the lifetime
effect for $B_s$ measurements at hadronic machines \cite{DescotesGenon:2011pb} for $B_s\to\mu\mu$~\cite{DeBruyn:2012wk} and $B_s\to\phi \mu\mu$~\cite{Descotes-Genon:2015hea} decays.

We implement the fit in two different ways.
First, we construct the full $\chi^2$ as a function of the model parameters, including all theoretical and experimental correlations, exactly as in Ref.~\cite{Descotes-Genon:2015uva}.\footnote{
The fit in Ref.~\cite{Descotes-Genon:2015uva} includes $b\to s\gamma$ observables. These observables are not included in our fit.
}
Second, in order to provide simplified expressions to allow the reader to repeat the fit without too much work, we perform a global fit to the relevant coefficients of the effective weak Hamiltonian 
 \eq{   
   \mathcal{H}_{\rm{eff}} \supset- \frac{  4 G_F }{\sqrt{2}}   \frac{\alpha}{4 \pi}   V_{ts}^*  V_{tb} \sum_{i=9,10} \left[   \C_{i \ell}   Q_{i}^{\ell} + \C_{i \ell}^{\prime}   Q_{i}^{\prime \ell}   \right] \,,
 }
with 
\begin{align}
\begin{aligned}
Q_{9}^{\ell} &= (  \bar s \gamma_{\alpha} P_L b  ) (  \bar \ell \gamma^{\alpha} \ell )  \,, \qquad Q_{9}^{\prime \ell} = (  \bar s \gamma_{\alpha}   P_R b  )( \bar \ell \gamma^{\alpha} \ell ) \,, \\
Q_{10}^{\ell} &= (  \bar s \gamma_{\alpha} P_L b  ) (  \bar \ell \gamma^{\alpha}  \gamma_5 \ell )  \,, \quad Q_{10}^{\prime \ell} = (  \bar s \gamma_{\alpha}   P_R b  )( \bar \ell \gamma^{\alpha}  \gamma_5 \ell ) \,.
\end{aligned}
\end{align}
We consider those coefficients receiving non-negligible NP contributions within our model, i.e. $(\C_{9\mu},\C_{10\mu},\C_{9e},\C_{10e})$, and provide the best fit points, standard deviations and
correlation matrix.\footnote{Contributions to the primed operators $Q_{9,10}^{\prime}$ are found to be negligible since the right-handed flavour changing $Z^{(\prime)}$ couplings to down-type quarks are suppressed by $m_f^2/u^2$, see \Sec{sec:strategy}.} These are collected in \Tab{tableflavor}.    The NP contributions to the Wilson coefficients ($\C_{i\ell}  = \C_{i\ell}^\text{SM} + \C_{i\ell}^\text{NP}$) are
\begin{align}    \label{eqwcsbtos}
\begin{aligned}
\C_{9a}^\text{NP}&=-\frac{\sqrt{2}}{G_F}\frac{\pi}{\alpha} \frac{1}{V_{tb}V_{ts}^*}\frac{\hat g^2}{8M_{W'}^2}(\Delta^q)_{bs}\left[(\Delta^\ell)_{aa}+\zeta  \left(4s_W^2-1\right)\right]  \,,\\[4mm]
\C_{10a}^\text{NP}&=\frac{\sqrt{2}}{G_F}\frac{\pi}{\alpha} \frac{1}{V_{tb}V_{ts}^*}\frac{\hat g^2}{8M_{W'}^2}(\Delta^q)_{bs}\left[(\Delta^\ell)_{aa}-\zeta  \right] \,.
\end{aligned}
\end{align}
Using these four coefficients
as ``pseudo observables" and constructing the $\chi^2$ function leads to a
linearised approximation to the fit. We have checked that the result of such a fit
is in reasonable agreement with the full fit.

\subsection{$b\to c$ transitions}

We consider the exclusive ratios
$R(D^{(*)})\equiv \Gamma(B\to D^{(*)}\tau \bar \nu)/\Gamma(B\to D^{(*)}\ell \bar \nu)$,
and the inclusive ratio $R(X_c)\equiv \Gamma(B\to X_c\tau \bar \nu)/\Gamma(B\to X_c\ell \bar \nu)$
as measures of flavour non-universality between the $\tau$ and the light leptons,
as well as the ratios $\Gamma(B\to D^{(*)}\mu \bar \nu)/\Gamma(B\to D^{(*)} e \bar \nu)$ constraining $e-\mu$
non-universality.  

The experimental value for the inclusive ratio $R(X_c)$ is obtained from the PDG averages for
$\mathrm{Br}(\bar b \rightarrow X \tau^+ \nu)$ and $\mathrm{Br}(\bar b \rightarrow X e^+ \nu)$.
The allowed size of lepton flavour universality violating effects in $b \rightarrow c \ell \nu$ ($\ell = e, \mu$) transitions
is not trivial to account for given that experimental analyses tend to present combined results for the
electron and muon data samples. This aspect was also stressed in Ref.~\cite{Greljo:2015mma}.  Experimental
results are however reported separately for the $e$ and $\mu$ samples in an analysis performed by the BaBar
collaboration~\cite{Aubert:2008yv}.  We use the values of  $\mathrm{Br}(B\to D^{(*)}\ell \bar \nu)$ reported in
Table IV of Ref.~\cite{Aubert:2008yv} to extract the ratios
$\Gamma(B\to D^{(*)}\mu \bar \nu)/\Gamma(B\to D^{(*)} e \bar \nu)$. The correlation between the two ratios is estimated from the information provided in~\cite{Aubert:2008yv}, adding the covariance for the systematic and statistical errors.
For the experimental values of $R(D)$ and $R(D^*)$ we consider the latest HFAG average~\hfag .  The latter includes $R(D)$ and $R(D^*)$ measurements performed by BaBar and Belle~\cite{Lees:2012xj,Huschle:2015rga},
the LHCb measurement of $R(D^*)$~\cite{Aaij:2015yra}, and the independent Belle measurement of $R(D^*)$ using a
semileptonic tagging method~\cite{Abdesselam:2016cgx}.\footnote{New results for $R(D^*)$ and the tau polarization asymmetry in $B \to D^* \tau \nu$ decays $(P_{\tau})$ using a hadronic tag have been presented by the Belle collaboration in Ref.~\cite{Abdesselam:2016xqt}.    The reported measurements are $R(D^*) = 0.276 \pm 0.034^{+0.029}_{-0.026}$ and $P_{\tau} = -0.44 \pm 0.47^{+0.20}_{-0.17}$~\cite{Abdesselam:2016xqt}. These measurements are not included in our analysis but would have a negligible impact if added given that the weighted average for $R(D^*)$ remains basically the same and the experimental uncertainty in $P_{\tau}$ is still very large.   Note that the measured tau polarization asymmetry is well compatible with the SM prediction $P_{\tau} = -0.502^{+0.006}_{-0.005} \pm 0.017$~\cite{Celis:2012dk}. } 
The results are summarized in \Tab{tableflavor}.

The model expressions for these ratios are:
\eqa{
\label{btoc1}
\frac{\Gamma(B^- \to  D^{(*)} \mu \bar \nu)}{\Gamma(B^- \to  D^{(*)} e \bar \nu)} &=&
\dfrac{\sum_j |\C_{2j}^{cb}|^2}{\sum_j |\C_{1j}^{cb}|^2}  \times 
\left[\frac{ \Gamma(B^- \to  D^{(*)} \mu \bar \nu)}{\Gamma(B^- \to  D^{(*)} e \bar \nu)}\right]_{\mbox{\scriptsize{SM}}} \ , \\[2mm]
\label{btoc2}
R(D^{(*)}) &=& \dfrac{2\,(\sum_j |\C_{3j}^{cb}|^2)}
{\sum_j (|\C_{1j}^{cb}|^2 + |\C_{2j}^{cb}|^2)} \times
R(D^{(*)})^{\mbox{\scriptsize{SM}}} \ , \\[2mm]
\label{btoc3}
R(X_c) &=&
\dfrac{\sum_j |\C_{3j}^{cb}|^2}{\sum_j |\C_{1j}^{cb}|^2} \times
R(X_c)^{\mbox{\scriptsize{SM}}} \ ,
}
where the Wilson coefficients $\C_{ij}^{cb}$ are given in \Eq{cijud}.  We use the SM predictions of $R(D)$ and $R(D^*)$ obtained in
Refs.~\cite{Kamenik:2008tj,Fajfer:2012vx}.
Note that recent determinations of $R(D)$ in Lattice QCD are compatible with the one used here~\cite{Lattice:2015rga,Na:2015kha}.
For $R(X_c)$ we use the SM prediction reported in Ref.~\cite{Ligeti:2014kia}.
For the ratios $\Gamma(B^- \to  D^{(*)} \mu \bar \nu)/\Gamma(B^- \to  D^{(*)} e \bar \nu)$ we derive the
SM predictions using the Caprini-Lellouch-Neubert parametrization of the form factors~\cite{Caprini:1997mu},
with the relevant parameters taken from HFAG~\hfag.
The resulting SM predictions are given in \Tab{tableflavor}.

\subsection{Lepton Flavour Violation}

We consider current limits on the lepton flavour violating decays $\tau \to 3 \mu$ and $Z \to \tau \mu$.    The decay $Z \to \tau \mu$ occurs due to gauge mixing effects.  The decay rate for $Z\to\tau\mu\equiv\left(\tau^+\mu^-+\tau^-\mu^+\right)$ is
\begin{align}
\Gamma\left(Z\to\tau\mu\right)=\frac{M_Z}{48\pi}\left(\zeta\, n_2\,\frac{g_2^4}{n_1^4}\Delta_\mu\Delta_\tau\,\ep2\right)^2\,.
\end{align}
We use the limit $\mathrm{Br}(Z \rightarrow \tau \mu)  <   1.2 \times 10^{-5}  $~\cite{Agashe:2014kda}.

The decay $\tau \to 3 \mu$ receives tree-level contributions from $Z^{(\prime)}$ exchange, the decay rate is given by
\begin{align}
\begin{aligned}
\Gamma\left(\tau\to3\mu\right)=\frac{\left[2\, (\mathcal{C}_{LL}^{\tau \mu})^2+   (\mathcal{C}_{LR}^{\tau \mu})^2 \right]m_\tau^5}{1536\pi^3} \,,
\end{aligned}
\end{align}
where the Wilson coefficients $\mathcal{C}_{LL}^{\tau \mu}$ and $\mathcal{C}_{LR}^{\tau \mu}$ are given by
%in \Eq{lfvwcs}.  
%
\begin{align}  \label{lfvwcs}
\begin{aligned}
\mathcal{C}_{LL}^{\ell_a \ell_b}&=\frac{\hat{g}^2}{4M_{W^\prime}^2} \Delta^\ell_{ab}   \left[  \Delta^\ell_{bb}  +\zeta \left(2s_W^2-1\right)\right]  \,, \\
\mathcal{C}_{LR}^{\ell_a \ell_b}&=\frac{\hat{g}^2}{2M_{W^\prime}^2}\,\zeta\,  \Delta^\ell_{ab} \, s_W^2 \,.
\end{aligned}
\end{align}
We use the HFAG limit $\mathrm{Br}(\tau \rightarrow 3 \mu )  <   1.2 \times 10^{-8} $~\hfag.

\section{Global fit}
\label{sec:fit}

\subsection{Fitting procedure}
We first fix the values of $g, g^{\prime}$ and the electroweak vev $v$ with the values of $\{ G_F,  \alpha,  M_Z \}$ reported in~\Tab{inputs}.   The $\mathrm{SU(2)}_1$ gauge coupling $g_1$ is then determined as a function of $g_2$.   The observables considered will depend on seven model parameters:
\begin{align*}
 M_{Z^{\prime}} &:  \text{The $Z^{\prime}$ mass, note that $M_{W^{\prime}} \simeq M_{Z^{\prime}}$} \,, \nonumber \\
   g_2 &:  \text{The $\mathrm{SU(2)}_2$ gauge coupling} \,, \nonumber \\
    \zeta &: \text{Controls the size of gauge mixing effects, see Eq.~\eqref{defmix}}  \,, \nonumber \\
   \Delta_s, \Delta_b, \Delta_{\mu}, \Delta_{\tau} &:    \text{Determine the gauge couplings to fermions, see Eq.~\eqref{eq:lambdamy}}   \,.  
 \end{align*}
The observables will also depend on the CKM inputs $\{ \lambda, A, \bar \rho, \bar \eta\}$.        We construct a global $\chi^2$ function that includes information from electroweak precision data at the $Z$ and $W$ poles together with flavour data.  It reads
\begin{align}
\chi^2 \equiv   (  O  - O_{\rm{exp}}     )^T  \overline \Sigma^{-1} (  O  - O_{\rm{exp}}     )     + \sum_{x=\lambda, A, \bar \rho, \bar \eta} \frac{(x- \hat x)^2}{\sigma_{\pm}^2} \,,
\end{align}
with $\overline \Sigma$ being the covariance matrix,  $O$ denoting the observables included in the analysis and $O_{\rm{exp}}$ the corresponding experimental mean values.   These are described in \Sec{sec:flavor}.  The CKM inputs $\{\lambda, A, \bar \rho , \bar \eta\}$ are included as pseudo-observables in the fit taking into account the values in~\Tab{inputs}.\footnote{These CKM inputs are obtained from a fit by the {\sc CKMfitter} group with only tree-level processes~\cite{CKMgroup}, as used in Ref.~\cite{Bazavov:2016nty}.}  The latter are reported in the form $\hat x^{+\sigma_+}_{-\sigma_-}$.  In the $\chi^2$ we introduce the asymmetric error: $\sigma_{\pm} = \sigma_+$ (for $x > \hat x$) and $\sigma_{\pm} = \sigma_-$ (for $x < \hat x$).     

\begin{table}[!tb]
\renewcommand{\arraystretch}{1.2}
\begin{centering}
\begin{tabular}{@{}c | c@{}}
\toprule[1.6pt] 
$\lambda = 0.22541(^{+30}_{-21})$     \hfill\cite{CKMgroup}   &   $A = 0.8212(^{+66}_{-338})$   \hfill\cite{CKMgroup}   \\
   $\bar \rho =   0.132(^{+21}_{-21}) $  \hfill\cite{CKMgroup}  &
$\bar \eta = 0.383(^{+22}_{-22})$  \hfill\cite{CKMgroup} \\    \midrule[0.6pt] 
  $G_F = 1.16638(1) \times 10^{-5}~\text{GeV}^{-2}$ \hfill\cite{Agashe:2014kda}
& $M_{Z} = 91.1876(21)~\text{GeV}$             \hfill\cite{Agashe:2014kda}
\\
  $\alpha = 1/137.036$         \quad         \hfill\cite{Agashe:2014kda}  &     \\
\bottomrule[1.6pt]
\end{tabular}  
\renewcommand{\arraystretch}{1.0}
  \caption{ \small \sf
    Electroweak and CKM inputs.
    }
  \label{inputs}
  \end{centering}
\end{table}

The global fit takes into account then seven model parameters $\{ M_{Z^{\prime}}, g_2, \Delta_s, \Delta_b, \Delta_{\mu}, \Delta_{\tau}, \zeta   \}$ and four CKM quantities $\{ \lambda, A, \bar \rho, \bar \eta\}$.      To sample the 11-dimensional parameter space we use the affine invariant Markov chain Monte Carlo ensemble sampler {\tt emcee}~\cite{ForemanMackey:2012ig}.

\subsection{Results of the fit}

We restrict the parameter space to $ 500~\text{GeV}  \leq M_{Z^{\prime}} \leq  3000~\text{GeV}$, $g < g_2 < \sqrt{4 \pi}$, $|\Delta_{a}| \leq 3 $ and $ 0 \leq  \zeta \leq 1 $.    The minimum of the $\chi^2$ is found to be at
\eq{  
\{ M_{Z^{\prime}}~\text{[GeV]}, g_2, \Delta_s, \Delta_b, |\Delta_{\mu}|, |\Delta_{\tau}|, \zeta \}
= \{1436, 1.04, -1.14, 0.016, 0.39, 0.075, 0.14\} \,,
}
with the CKM values $\{ \lambda, A, \bar \rho , \bar \eta \}$ within the $1 \sigma$ range in~\Tab{inputs}. 
It is enlightening to characterise the best-fit point in terms of the couplings appearing in the Lagrangian. We find that the corresponding Yukawas are, up to a global sign,
\begin{align}\label{eq:lBF}
\begin{aligned}
\lambda_\ell &\simeq\left(
  \begin{array}{cc}
 -1.2 & 0 \\
  0 & -0.3 \\
  0 & -0.06
  \end{array}
  \right) \, \times \frac{M_{L}}{\mathrm{TeV}}\,, \qquad 
\lambda_q &\simeq\left(
  \begin{array}{cc}
 -1.2 & 0 \\
  0 & 1.8 \\
  0 & -0.03
  \end{array}
  \right) \, \times \frac{M_{Q}}{\mathrm{TeV}}\,.
\end{aligned}
\end{align}

\begin{figure}[ht!]
\centering
\includegraphics[width=8cm]{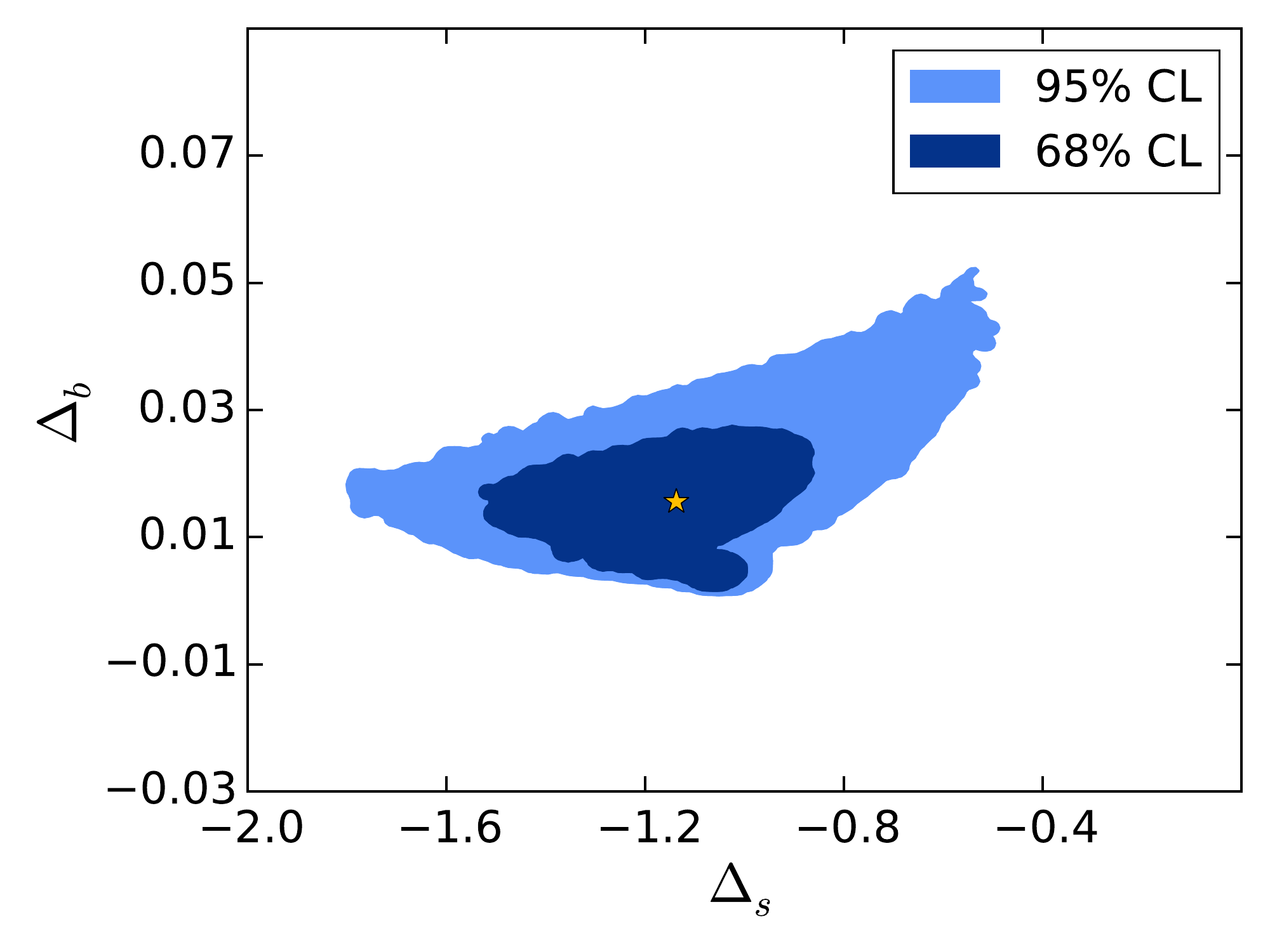}
~
\includegraphics[width=8cm]{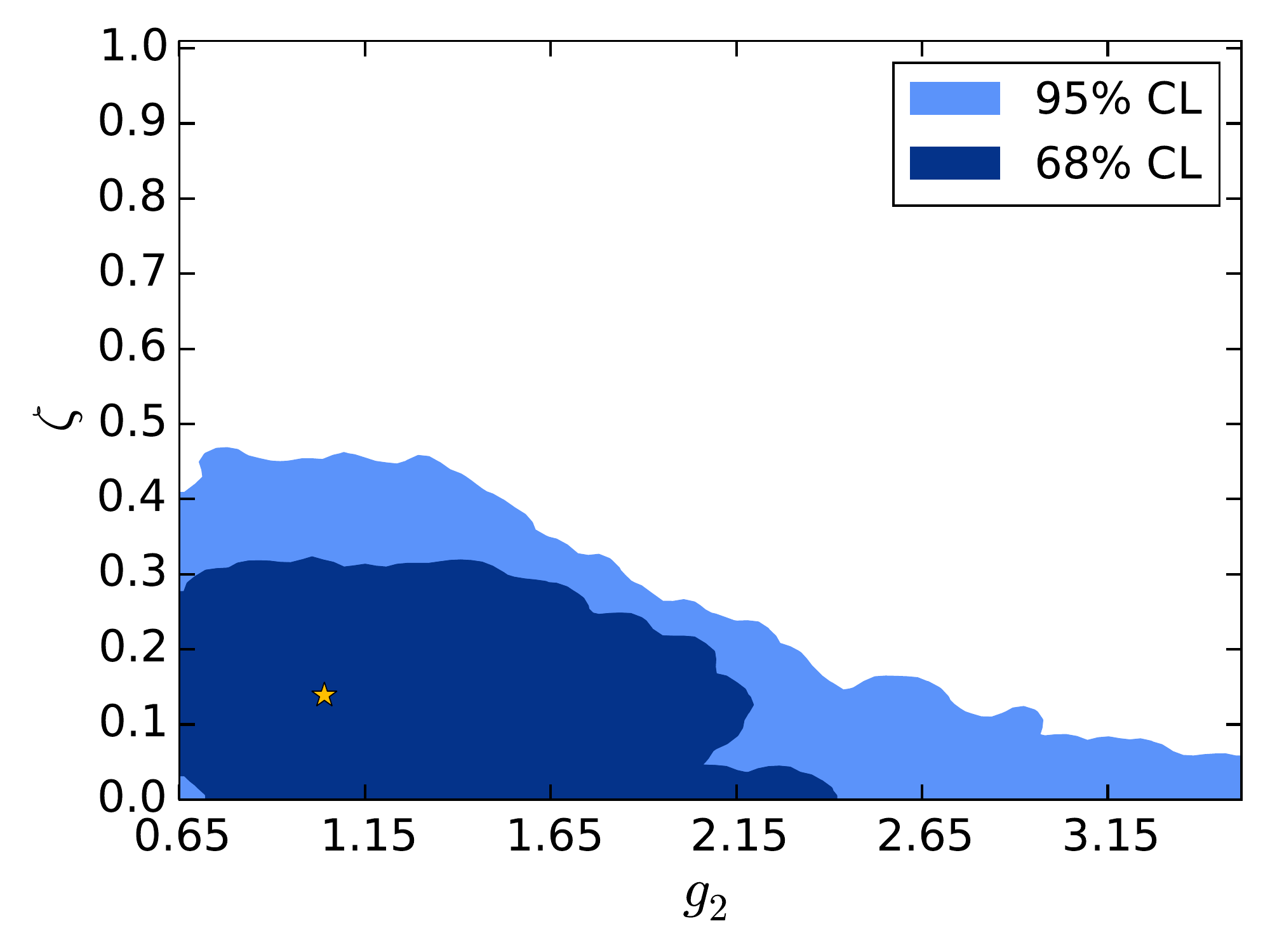}  \\
\includegraphics[width=8cm]{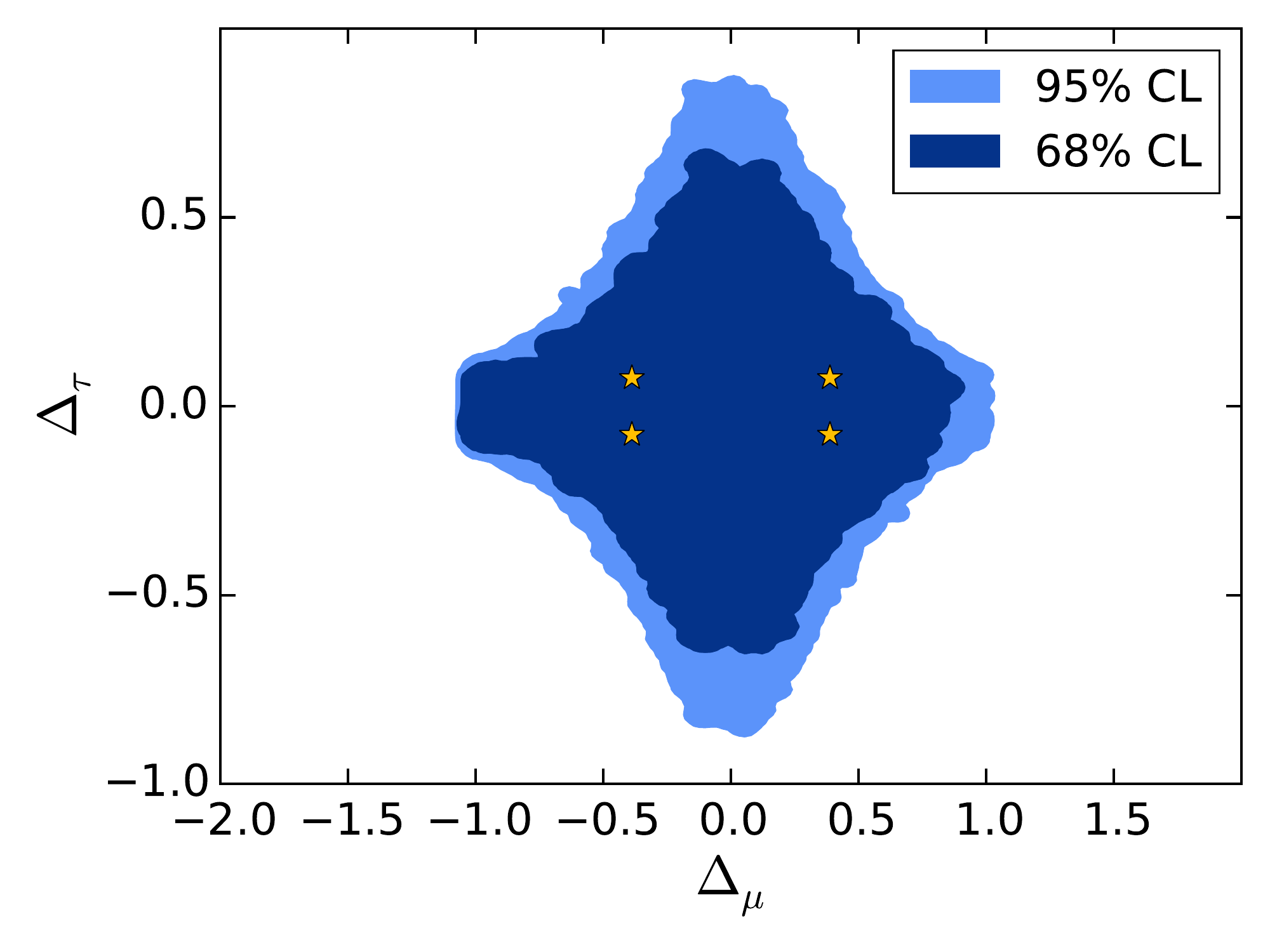} 
~
\includegraphics[width=8cm]{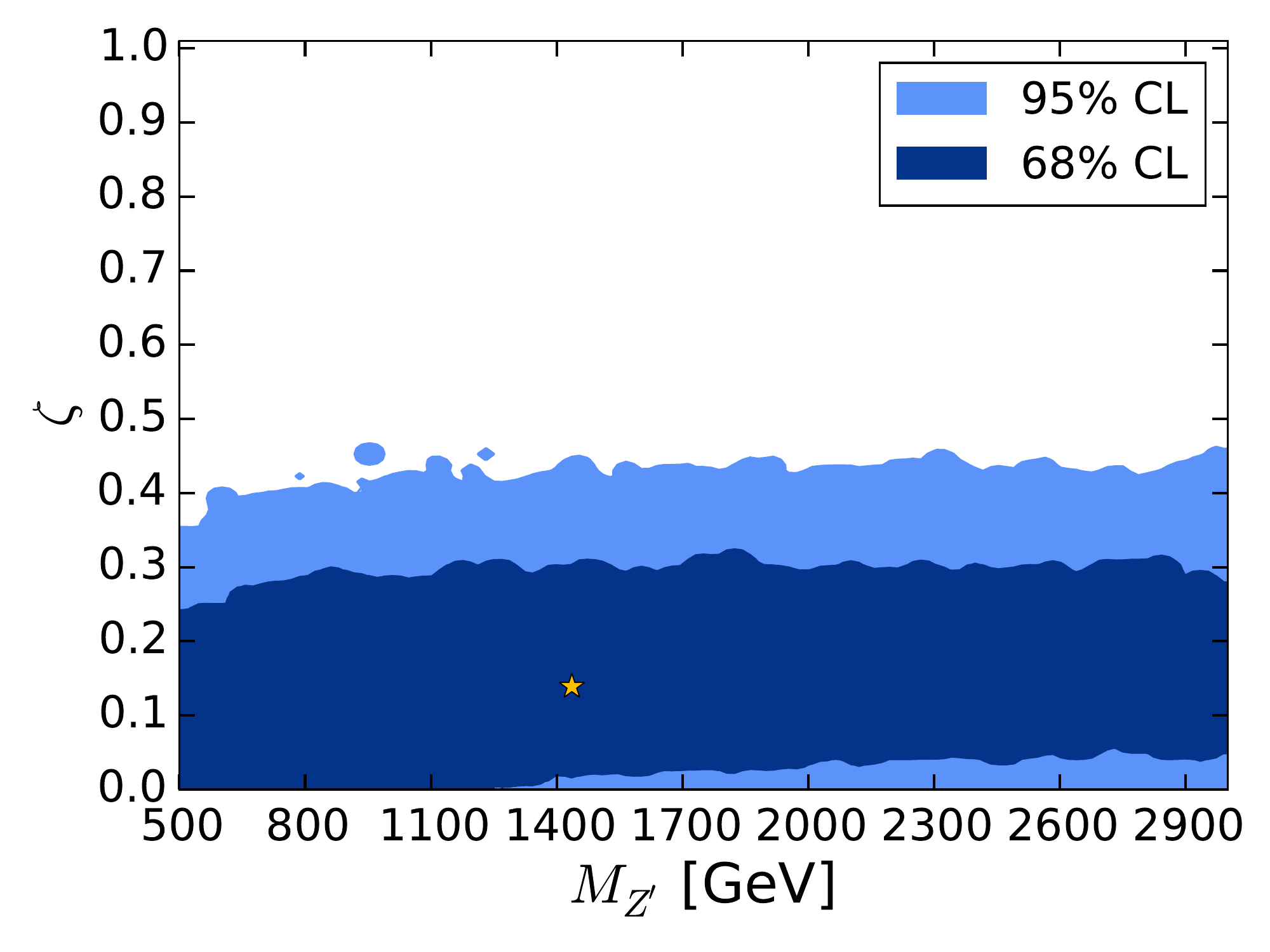}
\caption{\small \sf   Allowed regions for the model parameters at $68\%$ and $95\%$~CL from the global fit.    The best fit point is illustrated with a star.        }
\label{fig:phenopar}
\end{figure}

At the best-fit point we obtain $\chi^2_{\rm{min}} = 54.8$, to be compared with the corresponding value in the SM-limit $\chi^2_{\rm{SM}} = 93.7$.     We derive contours of $\Delta \chi^2 \equiv \chi^2 - \chi^2_{\rm{min}}$ in two-dimensional planes after profiling over all the other parameters, taking $\Delta \chi^2  = 2.3$ for $68 \%$~confidence level (CL) and $\Delta \chi^2  = 6.18$ for $95 \%$~CL.    Allowed regions for the model parameters obtained in this way are shown in Figure~\ref{fig:phenopar}.     

%%%%%%%%%%%%%%%%%%%%%%%%%%%%%%%%%%%%%%%%%%%%%%%%%%%%%%%%%%%%%%%%%%%%%%%%%
%%%%%%%%%%%%%%%%%%%%%%%%%%%%%%%%%%%%%%%%%%%%%%%%%%%%%%%%%%%%%%%%%%%%%%%%%%

\begin{figure}[ht!]
\centering
\includegraphics[width=8.0cm]{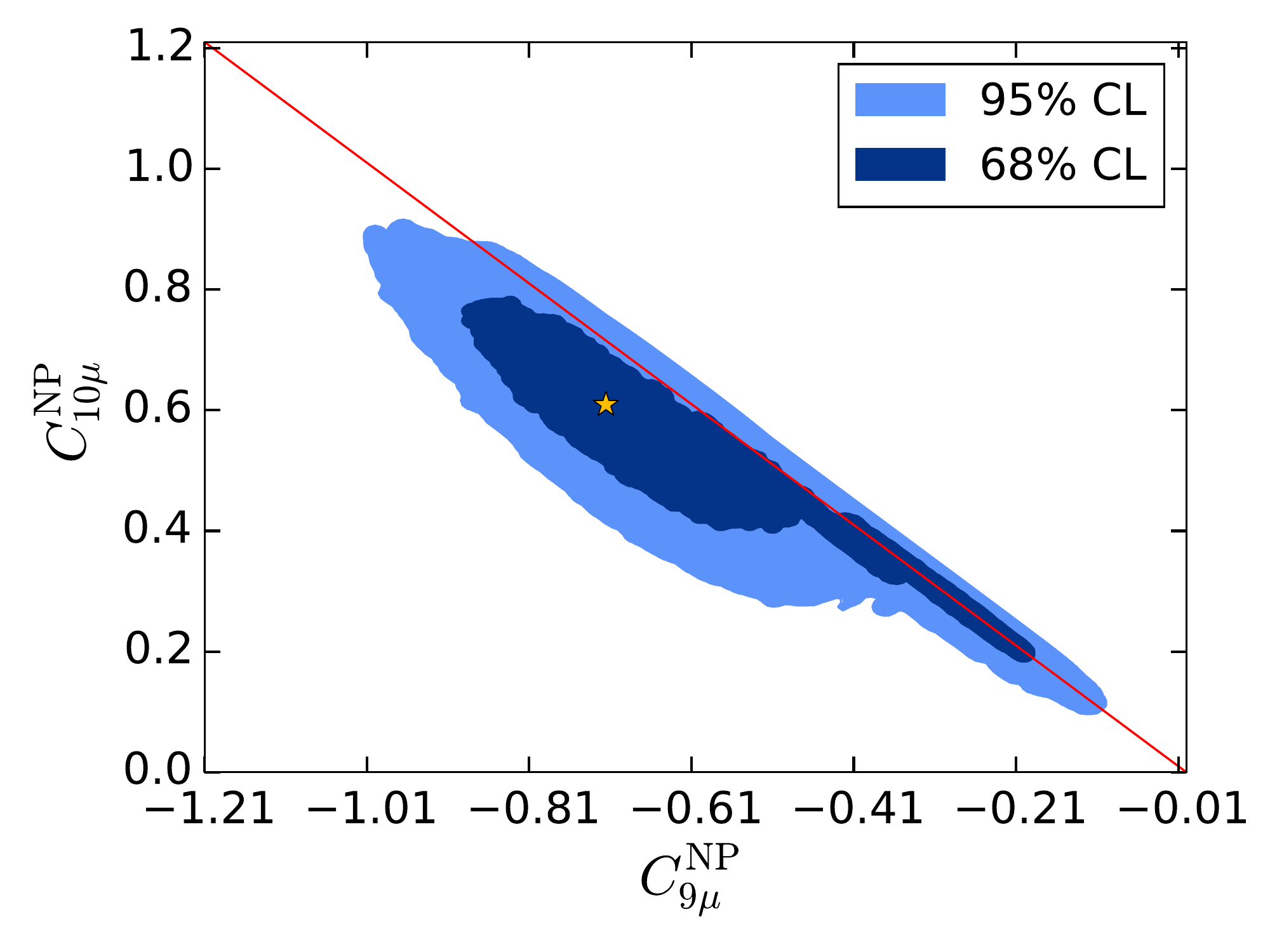}
~
\includegraphics[width=8.0cm]{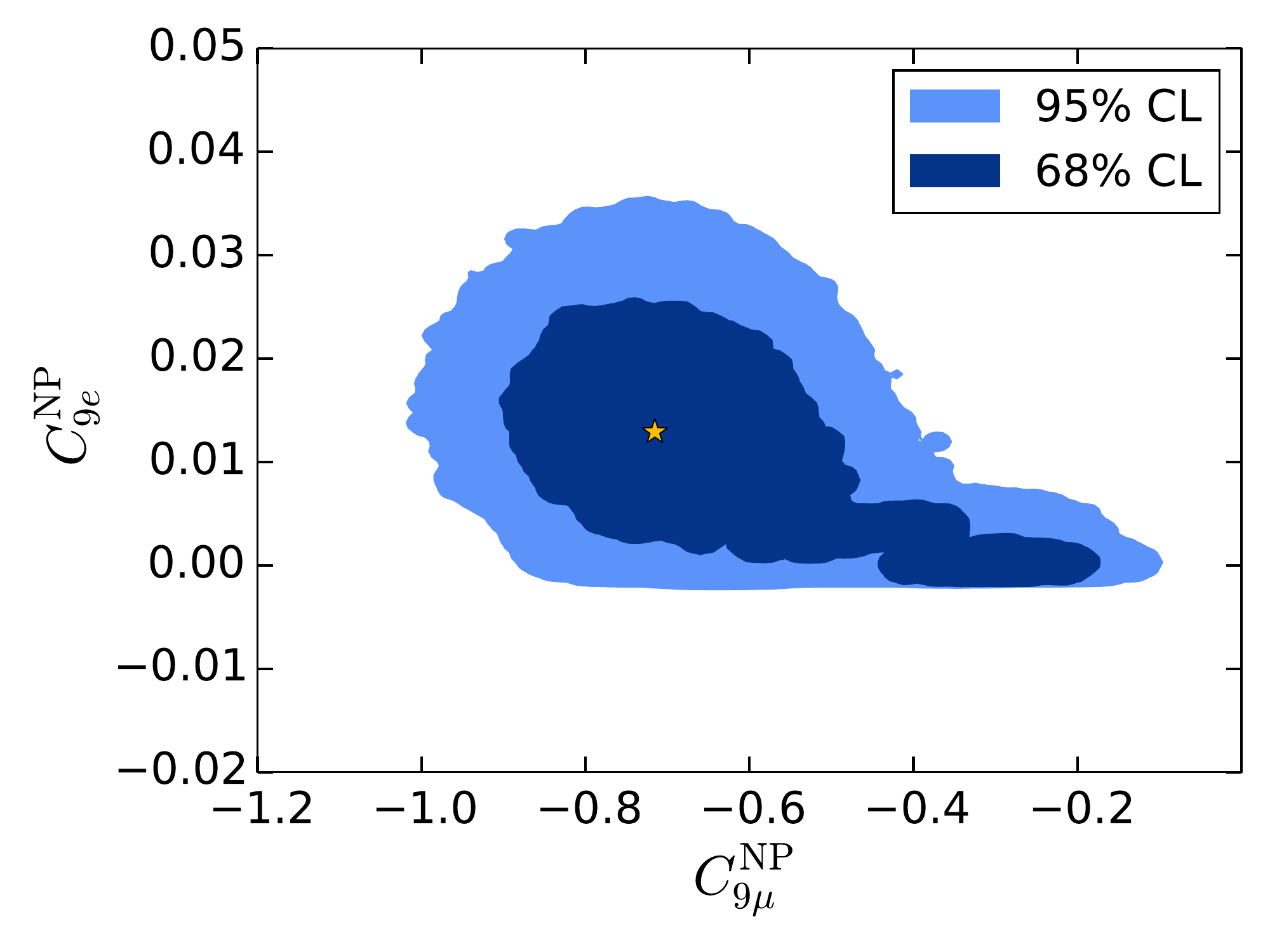} 
\caption{\small \sf   Allowed regions at $68\%$ and $95\%$~CL from the global fit for the Wilson coefficients of $b \to s \ell^+ \ell^-$ transitions.    The best fit point is illustrated with a star.  The red line on the left plot illustrates the correlation $\C_{9\mu}^\text{NP}  = - \C_{10\mu}^\text{NP}$.      }
\label{fig:phenoIb}
\end{figure}

There is a four-fold degeneracy of the $\chi^2$ minimum  with the sign of $\Delta_{\mu, \tau}$ as no observable in the fit is sensitive to the relative sign between $\Delta_{\mu}$ and $\Delta_{\tau}$.     The allowed values of $\Delta_{\mu, \tau}$ lie in the region $|\Delta_{\mu, \tau}| \lesssim 1$.     While $\Delta_b$ is bounded to be very small $\sim 10^{-2}$, the allowed values for $\Delta_s$ are around $-1$.  The negative sign obtained for the combination $\Delta_s \Delta_b$ is related to the preference for negative values of $\C_{9\mu}^\text{NP}$ by $b \to s \ell^+ \ell^-$ data.          The allowed regions for the Wilson coefficients of $b \to s \ell^+ \ell^-$ transitions from the global fit are shown in Figure~\ref{fig:phenoIb}.    Note that with the assumed flavour structure we have the correlation $\C_{10e}^\text{NP}  =  (4 s_W^2 -1) \C_{9e}^\text{NP}$.  The relation $\C_{9\mu}^\text{NP}  = - \C_{10\mu}^\text{NP}$ on the other hand holds in our model only in the absence of gauge mixing effects.  Departures from this correlation are possible as gauge mixing effects can be sizeable, see Figure~\ref{fig:phenoIb} (left).

%%%%%%%%%%%%%%%%%%%%%%%%%%%%%%%%%%%%%%%%%%%%%%%%%%%%%%%%%%%%%%%%%%%%%%%%%
%%%%%%%%%%%%%%%%%%%%%%%%%%%%%%%%%%%%%%%%%%%%%%%%%%%%%%%%%%%%%%%%%%%%%%%%%%

\begin{figure}[ht!]
\centering
\includegraphics[width=8.0cm]{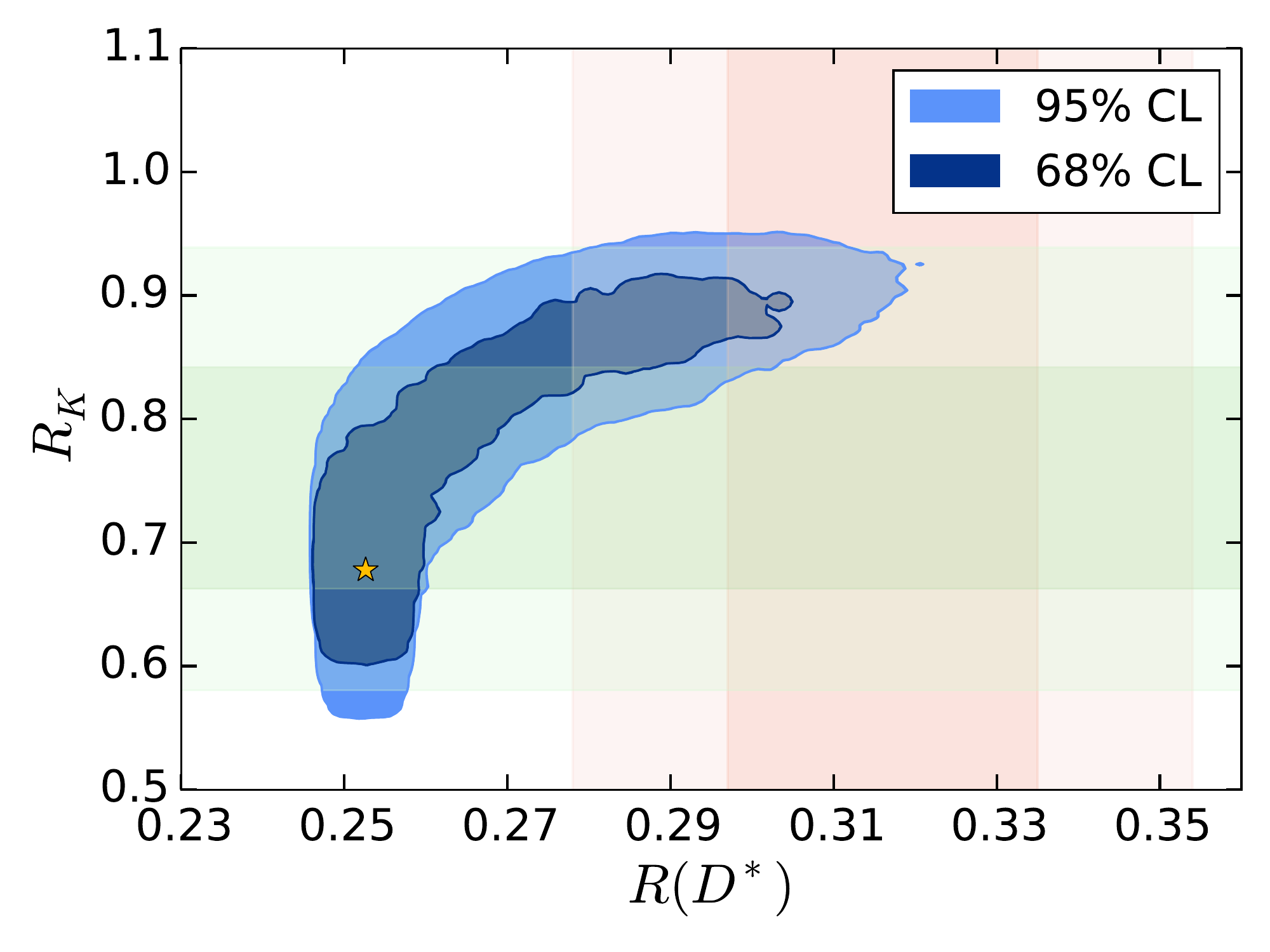} 
~
\includegraphics[width=8.0cm]{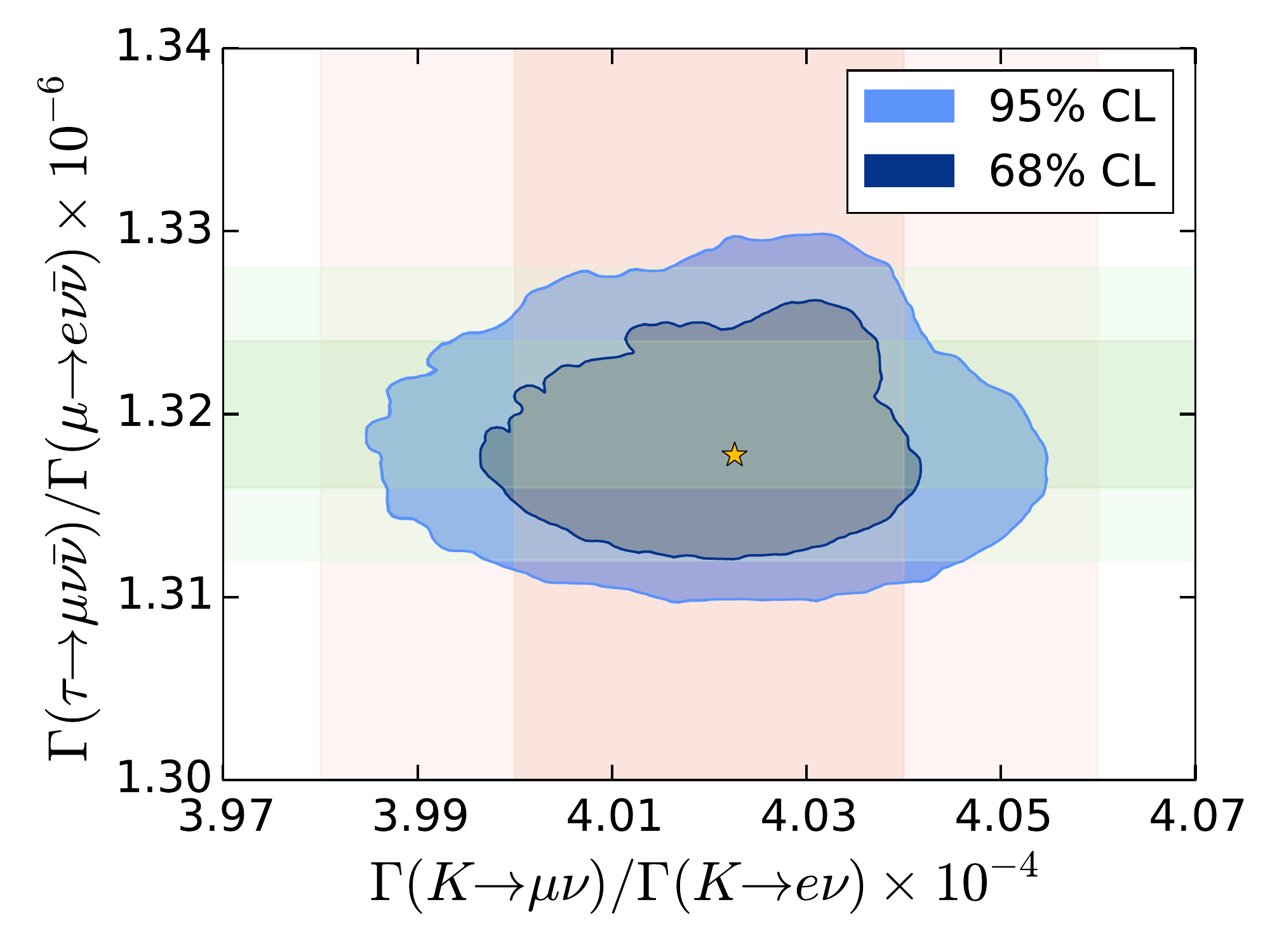} 
\\
\includegraphics[width=8.0cm]{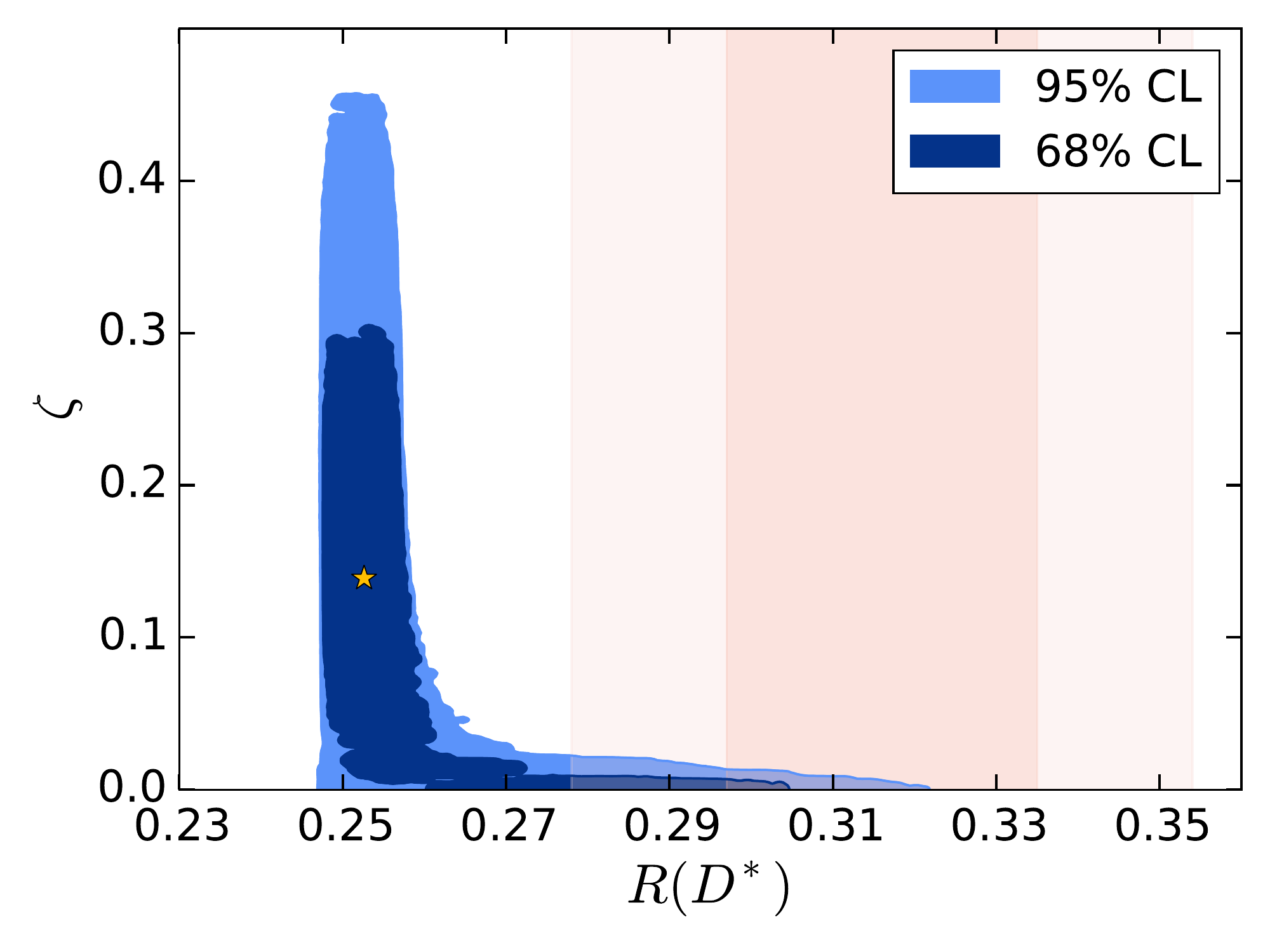}
~
\includegraphics[width=8.0cm]{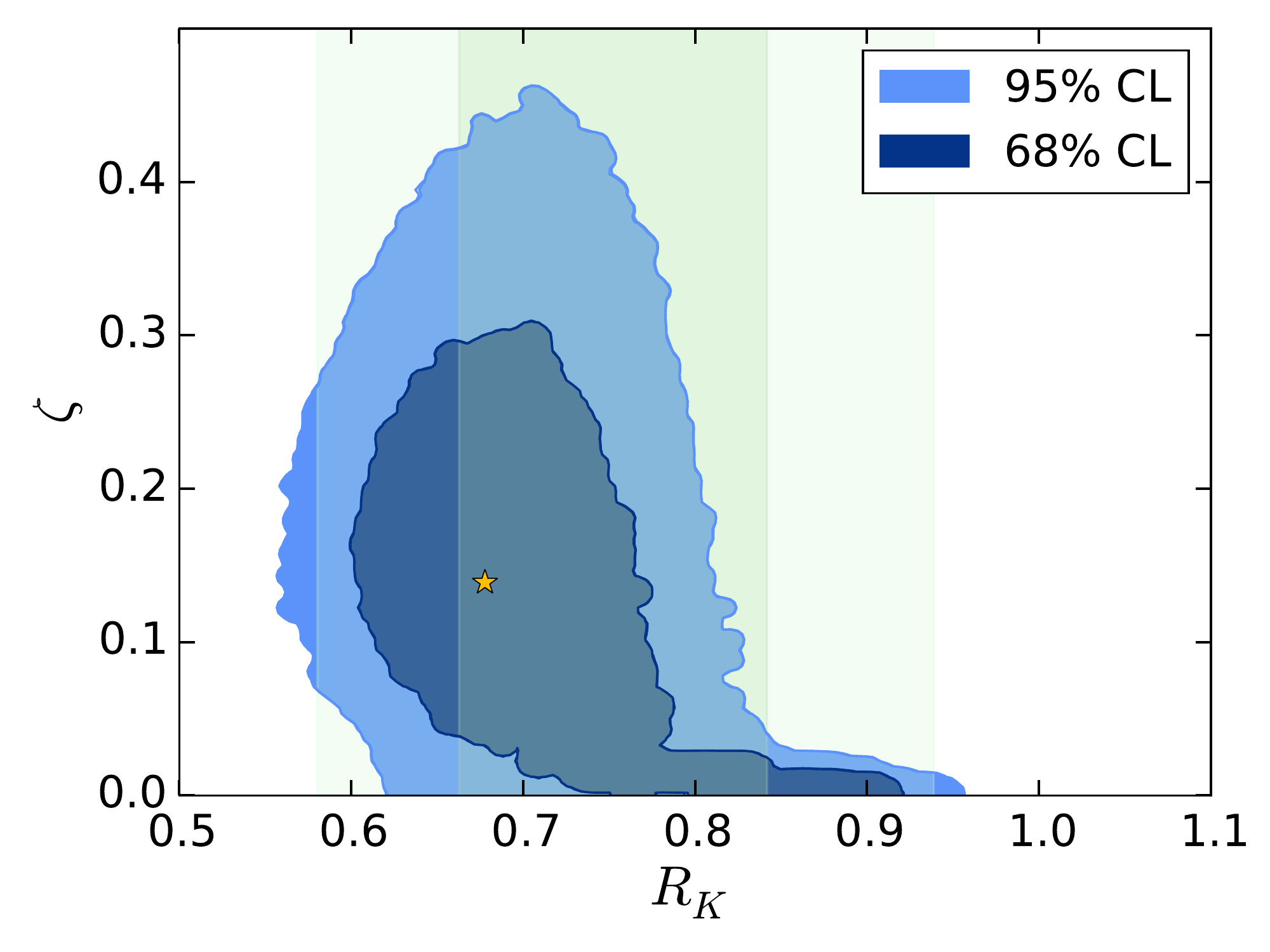} 
\caption{\small \sf   Allowed regions at $68\%$ and $95\%$~CL from the global fit.    Experimental values for these observables are also shown at $1\sigma$ (dark-band) and $2 \sigma$ (light-band).   The best fit point is illustrated with a star.        }
\label{fig:phenoI}
\end{figure}

Allowed values at $68\%$ and $95\%$~CL for $R_K$ and $R(D^{*})$ are shown in Figure~\ref{fig:phenoI}.   The best fit point presents a sizeable deviation from the SM in $R_K$ in the direction of the LHCb measurement while the ratios $R(D^{*})$ are SM-like.      Note that the NP scaling of $R(D)$ is the same as for $R(D^*)$ because the $W^{\prime}$ couplings are mostly left-handed, with the right-handed couplings suppressed by $m_f^2/u^2$.
A significant enhancement of $R(D^{(*)})$ is possible within the allowed parameter region.  The model presents a positive correlation between $R_K$ and $R(D^{(*)})$ so that $R_K$ is above its best-fit value whenever $R(D^{(*)})$ gets enhanced.    The ratios $\Gamma(B \to D^{(*)} \mu \nu)/\Gamma(B \to D^{(*)} e \nu)$ are found to be SM-like with possible deviations only at the $\sim1\%$ level.    As expected,  $R(X_c)$ and $R(D^{(*)})$ show a strong correlation, in the region of the parameter space where $R(D^{(*)})$ accommodates the current experimental values one obtains a slight tension in  $R(X_c)$ with experiment.      The flavour observables with light-mesons and leptonic $\tau$-decays are found to be in good agreement with the SM and experiment, we show the resulting allowed values for $K \to \mu \nu/K \to e \nu$ and $\tau \to \mu \nu \bar \nu/\mu \to e \nu  \bar \nu$ as an example in Figure~\ref{fig:phenoI}.

%%%%%%%%%%%%%%%%%%%%%%%%%%%%%%%%%%%%%%%%%%%%%%%%%%%%%%%%%%%%%%%%%%%%%%%%%
%%%%%%%%%%%%%%%%%%%%%%%%%%%%%%%%%%%%%%%%%%%%%%%%%%%%%%%%%%%%%%%%%%%%%%%%%%

As noted in Ref.~\cite{Boucenna:2016wpr}, gauge mixing effects play a crucial role in the possible enhancement of $R(D^{(*)})$ in this model.    In Figure~\ref{fig:phenoI} we also show the results of the global fit for $R(D^*)$ as a function of the parameter controlling the size of gauge mixing effects $\zeta$.     Having an enhancement of $R(D^{*})$ of order $\sim 20\%$ as suggested by the experimental measurements is only possible for $\zeta\ll1$.  The situation is very different for $R_K$, with the parameter $\zeta$ playing no major role in this case as shown in Figure~\ref{fig:phenoI}.    We find that the allowed points from the global fit accommodating both $R_K$ and $R(D^{(*)})$ within $2 \sigma$ lie within a very restricted region:
\begin{align}    \label{exkfn}
M_{Z^{\prime}} \in&\; [500, 1710]~\text{GeV} \,, \quad g_2 \in [1.2, 3.5] \,, \quad \Delta_s \in [-1.16, -0.97] \,, \quad \Delta_b \in [0.003, 0.007] \,, \nonumber \\[0.2cm]
|\Delta_{\mu}| \in&\;  [0.94,0.99] \,, \quad \quad  \,  |\Delta_{\tau}| \in [0, 0.11] \,,  \,~~\quad \zeta \in [0, 0.02] \,. 
\end{align}
The $Z^{\prime}$ mass and the $\mathrm{SU(2)}_2$ gauge coupling $g_2$ are positively correlated, going from $g_2 \sim 1$ for $M_{Z^{\prime}} \sim 500$~GeV up to the perturbativity limit $g_2 \leq \sqrt{4 \pi}$ for $M_{Z^{\prime}} \sim 1700$~GeV.  A limit on $\tan \beta$ can be derived in this region using \Eq{defmix}, we get $\tan \beta \in [0.2, 0.65]$.   Similarly, in this region the $\mathrm{SU(2)}_1$ gauge coupling satisfies $0.66 \leq g_1 \leq 0.78$ and the combination $\hat g = g g_2/g_1$ is found to be within $1 \leq \hat g \leq 3.4$.  Note that the $Z^{\prime}$ and $W^{\prime}$ interactions with the SM fermions are proportional to $1 - \Delta_{a}^2$, see \Eq{eq:lambdamy}.    In the parameter space region where both $R_K$ and $R(D^{(*)})$ are accommodated within $2 \sigma$, the massive gauge bosons, $Z^{\prime}$ and $W^{\prime}$, couple predominantly to the third fermion generation.

\section{Predictions}
\label{sec:predictions}

In the following we take the current measured values of $R_K$ and $R(D^{(*)})$ at face value, focusing on the parameter space region described in \Eq{exkfn}.  We are interested in possible signatures that can be used to test or falsify this scenario with upcoming measurements at the LHC and flavour factories.

\subsection{Differential distributions in $B \to D^{(*)} \tau \nu$ decays}

Due to the gauge structure of the model, new physics contributions to the $B \to D^{(*)} \ell \nu$ decay amplitudes have the same Dirac structure as the SM contribution to a good approximation.  This gives rise to a clean prediction 
\begin{align}  \label{eqkfnf}
\frac{  R(D) }{  R(D^*)  } = \left[ \frac{  R(D) }{  R(D^*)  } \right]_{\rm{SM}} \,,
\end{align}
which is compatible with current data~\cite{Amhis:2014hma}.    The inclusive ratio $R(X_c)$ can provide an additional handle to test the proposed scenario.   The model gives rise to an enhancement in  $R(X_c)$ within the parameter space region considered, we obtain $0.24 \leq R(X_c) \leq 0.29$.          The Dirac structure of the new physics contributions can also be tested by using information from the $q^2 \equiv (p_B - p_{D^{(*)}})^2$ spectra and by measuring additional observables that exploit the rich kinematics and spin of the final state particles.  The differential decay rate for $B \to D^{(*)} \tau \nu$ is affected in the model with a global rescaling factor, implying that forward-backward asymmetries as well as the $\tau$ and $D^*$ polarization fractions are expected to be as in the SM.     For recent studies of differential distributions in $b \to c \tau \nu$ decays see Refs.~\cite{Fajfer:2012vx,Datta:2012qk,Sakaki:2012ft,Celis:2012dk,Biancofiore:2013ki,Sakaki:2014sea,Duraisamy:2014sna,Shivashankara:2015cta,Bhattacharya:2015ida,Becirevic:2016hea,Sahoo:2016nvx,Alonso:2016gym,Alok:2016qyh,Bernlochner:2016bci}.        Future measurements of $b \to c \tau \nu$ transitions at the Belle-II experiment will be crucial to disentangle possible new physics contributions in these decays~\cite{Abe:2010gxa}.

\subsection{Lepton universality tests in $R_M$}

Confirming the violation of lepton flavour universality in other $b \to
s$ observables would be definite evidence in favour of new physics at
work. Examples of such additional observables are $R_M$, with $M = K^*, \phi$
\cite{Hiller:2014ula,Altmannshofer:2015mqa}, defined analogously to
$R_K$,
\begin{equation}
R_M[q^2_1,q^2_2] = \frac{ \int_{q^2_1}^{q^2_2} dq^2 \ d\Gamma(B_q \rightarrow M \mu^+ \mu^-)/dq^2}{\int_{q^2_1}^{q^2_2} dq^2\ d\Gamma(B_q \rightarrow M e^+ e^-)/dq^2}  \ ,
\end{equation}
with $q=d,s$ for $M = K^*, \phi$.\footnote{See Ref.~\cite{Capdevila:2016ivx} for other observables in $B\to K^* \ell\ell$ testing lepton-flavour non-universality.}

The expected values for $R_K$, $R_{K^*}$ and $R_\phi$ within each bin are strongly
correlated, except for the fact that hadronic uncertainties are mostly independent (but small). From the results of the fit, we find the following expected ranges for the different
ratios:
\begin{equation}
\begin{array}{lcllcl}
R_K[1,6] &\in& [0.62, 0.91] \text{ at } 68\% \text{ CL}
\ , \qquad &
R_K[1,6] &\in& [0.57, 0.95]  \text{ at } 95\% \text{ CL}\ ,
\\[2mm]
R_{K^*}[1.1,6] &\in& [0.66, 0.91]  \text{ at } 68\% \text{ CL}
\ , \qquad &
R_{K^*}[1.1,6] &\in& [0.62, 0.95]  \text{ at } 95\% \text{ CL}\ ,
\\[2mm]
R_{K^*}[15,19] &\in& [0.61, 0.90]  \text{ at } 68\% \text{ CL}
\ , \qquad &
R_{K^*}[15,19] &\in& [0.56, 0.94]  \text{ at } 95\% \text{ CL}\ ,
\\[2mm]
R_\phi[1.1,6] &\in& [0.64, 0.91]  \text{ at } 68\% \text{ CL}
\ , \qquad &
R_\phi[1.1,6] &\in& [0.60, 0.94]  \text{ at } 95\% \text{ CL}\ ,
\\[2mm]
R_\phi[15,19] &\in& [0.61, 0.90]  \text{ at } 68\% \text{ CL}
\ , \qquad &
R_\phi[15,19] &\in& [0.56, 0.94]  \text{ at } 95\% \text{ CL}\ ,
\end{array}
\end{equation}
where it is understood that a strong (positive) correlation exists among all the predictions,
lower values of one observable corresponding to lower values of another and viceversa.

\subsection{Lepton flavour violation} 

One of the first \emph{generic} consequences of the violation of lepton flavour universality is lepton flavour violation \cite{Glashow:2014iga}, as explored in connection to the $B$-meson anomalies in Refs. \cite{Boucenna:2015raa,Crivellin:2015era,Calibbi:2015kma,Guadagnoli:2015nra,Sahoo:2015pzk,Crivellin:2016vjc,Becirevic:2016zri,Kumar:2016omp,Guadagnoli:2016erb,Kim:2016bdu,Sahoo:2016nvx}.    In our model, the branching fraction for $\tau \to 3 \mu$ is proportional to $\Delta_{\tau}^2$ and is therefore suppressed for $|\Delta_{\tau}| \simeq 0$.   When $|\Delta_{\tau}|$ is near its upper bound, $|\Delta_{\tau}| \simeq 0.1$, we obtain values for $\mathrm{Br}(\tau \to 3 \mu)$ that saturate the current experimental limit $1.2 \times10^{-8}$.    Semileptonic decays of the tau lepton into a muon and a pseudo-scalar meson also receive tree-level contributions from $Z^{(\prime)}$ exchange, these will also be proportional to $\Delta_{\tau}^2$ so that the largest rates possible will be obtained for $|\Delta_{\tau}| \simeq 0.1$.     In our model the decays $\tau \to \mu \eta^{(\prime)}$ receive important new physics contributions through the axial-vector strange-quark current.   Following \cite{Celis:2014asa} we obtain $\mathrm{Br}(\tau \to \mu \eta^{\prime}) \leq 3.9 \times 10^{-8}$ and $\mathrm{Br}(\tau \to \mu \eta) \leq 4.2 \times 10^{-8}$, very close to the current experimental limits $\mathrm{Br}(\tau \to \mu \eta^{\prime})_{\rm{exp}} \leq 1.3 \times 10^{-7}$ and $\mathrm{Br}(\tau \to \mu \eta)_{\rm{exp}} \leq 6.5 \times 10^{-8}$~\cite{Miyazaki:2007jp}.       The observation of lepton flavour violating tau decays decays might therefore lie within the reach of future machines such as Belle-II, where an improvement of the current experimental bounds by an order of magnitude can be expected~\cite{Abe:2010gxa}.   On the other hand, due to the suppression of gauge mixing effects ($\zeta \ll 1$) the decay $Z \to \tau \mu$ lies well-below the current experimental limit, for which we obtain $\mathrm{Br}(Z \to  \mu \tau)  \leq 1.2 \times 10^{-9}$.

\subsection{Direct searches for new states at the LHC}

In this model we expect a plethora of new states lying at the TeV scale: scalar bosons (in the CP-conserving limit we would have two CP-even Higgs bosons, one CP-odd Higgs and one charged scalar, cf. Section~\ref{sec:model}), heavy fermions and the massive vector bosons $W^{\prime}$, $Z^{\prime}$.

The heavy vector-like leptons will be pair-produced at the LHC via Drell-Yan processes due to their coupling to the massive electroweak gauge bosons.   These will decay into gauge bosons and charged leptons or neutrinos.        Though no dedicated searches for vector-like leptons have been performed at the LHC, one can obtain limits on their mass and production cross-section by recasting existing multilepton searches~\cite{Falkowski:2013jya}.    It was found that current limits for a heavy lepton doublet decaying to $\ell=e,\mu$ flavours are around $450$~GeV while for decays into $\tau$-leptons the limits are around $270$~GeV~\cite{Falkowski:2013jya}.  Searches for pair production of heavy vector-like quarks at the LHC focus primarily into final states with a third generation fermion and bosonic states, setting upper limits on the vector-like quark masses ranging from $\sim700$~GeV up to $\sim1$~TeV~\cite{Aad:2014efa,Aad:2015kqa,Aad:2015gdg,Aad:2016qpo,Khachatryan:2015axa,Khachatryan:2015oba}.

The massive vector bosons $W^{\prime}$, $Z^{\prime}$ couple predominantly to the third fermion generation.     The LHC phenomenology of this type of states has been discussed in Ref.~\cite{Greljo:2015mma}.       The $Z^{\prime}$ coupling to muons is found to be at most $\sim12\%$ of its coupling to $\tau$-leptons.  In the quark sector, the $Z^{\prime}$ coupling to the second quark generation is found to be at most $\sim36\%$ of the coupling to third generation quarks.              The $Z^{\prime}$ boson would be produced at the LHC via Drell-Yan processes due to its coupling to $b$-quarks and $s/c$-quarks.

The total $Z^{\prime}$ width normalized by the $Z^{\prime}$ mass $(\Gamma_{Z^{\prime}}/M_{Z^{\prime}})$ is found to grow with $M_{Z^{\prime}}$, since $\hat{g}$ and $M_{Z^\prime}$ are positively correlated.    Assuming that the $Z^{\prime}$ can only decay into the SM fermions we have 
\begin{align}
\frac{\Gamma_{Z^{\prime}}}{M_{Z^{\prime}}} \simeq    \frac{  \hat g^2}{48 \pi} \left[ 3\sum_{q=  s,b }   (1 - \Delta_q^2)^2  + \sum_{\ell =  \mu, \tau }      (1 - \Delta_{\ell}^2)^2 \right] \,,
\end{align}
where we have neglected fermion mass effects.   We obtain that $\Gamma_{Z^{\prime}}/M_{Z^{\prime}}$ is between $2\%$ and $31\%$, with $\Gamma_{Z^{\prime}}/M_{Z^{\prime}} \gtrsim 10\%$ for $M_{Z^{\prime}} \gtrsim 1$~TeV.

If kinematically open, additional decay channels of the $Z^{\prime}$ boson would reduce the branching fractions to SM particles by enhancing the total $Z^{\prime}$ width, making the $Z^{\prime}$ resonance broader.     The latter scenario will generically be the case provided the vector-like fermions are light enough, opening decay channels of the $Z^{\prime}$ boson into a heavy vector-like fermion and a SM-like fermion or into a vector-like fermion pair.    The decay rate for these processes is given by:
\begin{align}
\begin{aligned}
\Gamma(Z^{\prime} \to F_i \bar f_j) &\simeq    \frac{  \lambda^{1/2}(1,x_i,x_j) \, \hat g^2  N_C   M_{Z^{\prime}}     }{192 \pi}   \left[ 2 - x_i  - x_j  - (  x_i - x_j  )^2      \right] (  \Sigma_{ij} )^2      \,, \\
\Gamma(Z^{\prime} \to F_i \bar F_i) &\simeq  \frac{  \lambda^{1/2}(1,x_i,x_i)   \, \hat g^2  N_C   M_{Z^{\prime}}      }{96 \pi} \left\{  (  1  -   x_i  )   \left(   (\Omega^{Q,L}_{ii})^2 + \frac{g_1^4}{g_2^4}        \right)  - 6\, \frac{g_1^2}{g_2^2}\,  x_i   \,  \Omega^{Q,L}_{ii}     \right\}  \,.
\end{aligned}
\end{align}
Here $\lambda(x,y,z) = x^2 + y^2 + z^2 -2 (x y + yz + x z)$,  $N_C = 3(1)$ for (un)coloured fermions and $x_i = m_i^2/M_{Z^{\prime}}^2$.  We have denoted by $F_i$ a generic heavy fermion and by $f_j$ one of the SM-like fermions.    The matrices $\Sigma$ and $\Omega^{Q,L}$ have been defined in Eqs.~\eqref{eq:lambdamy2} and \eqref{eq:lambdamy3}.   The $Z^{\prime}$ decays into a heavy fermion and a SM-like fermion are accidentally suppressed due to the small entries of the $\Sigma$ matrix within the parameter  region of interest.  These decays therefore give small contributions to the total width in general.    The decays into a pair of heavy fermions, on the other hand, can give a significant contribution to the total $Z^{\prime}$ width when kinematically allowed.   For instance, if the masses of the heavy leptons lie around $450$~GeV we obtain a contribution to $\Gamma_{Z^{\prime}}/M_{Z^{\prime}}$ from the decays $Z^{\prime} \to E_i \bar E_i, N_i \bar N_i$ ($i=1,2$) of about $20\%$ for $M_{Z^{\prime}} \sim 1.2$~TeV, making the $Z^{\prime}$ boson a very wide resonance in this case: $\Gamma_{Z^{\prime}}/M_{Z^{\prime}} \sim 30\%-50 \%$.

The ATLAS and CMS collaborations have searched for a resonance in the $\tau^+ \tau^-$ channel at $\sqrt{s} = 8$~TeV~\cite{Aad:2015osa,Aad:2014cka,CMS:2015ufa,Khachatryan:2014fba}.  Among these, the strongest limits are those coming from ATLAS and they place important bounds on the model.    We have evaluated the $Z^{\prime}$ production cross-section at the LHC using~{\tt MadGraph (MG5\textunderscore aMC\textunderscore 2.4.2)}~\cite{Alwall:2014hca}. We find that it is possible to exclude the low-mass region where the $Z^{\prime}$ resonance remains reasonably narrow and there is not much room for additional decay channels giving large contributions to the total width. The latter would require having very light exotic fermions, entering in conflict with direct searches for these states at colliders.   In the heavy $Z^{\prime}$ mass region ($\gtrsim 1$~TeV) the $Z^{\prime}$ resonance becomes wide $(\Gamma_{Z^{\prime}}/M_{Z^{\prime}} \gtrsim 10\%)$ and the interpretation of the current experimental results based on the search of a relatively narrow resonance is not valid anymore. Dedicated searches at the LHC for a broad resonance in the $\tau^+ \tau^-$ channel within the mass range $\sim1-1.7$~TeV would then be needed in order to test this scenario.\footnote{We find our main conclusions in this regard to agree with those posed previously by the authors of Ref.~\cite{Greljo:2015mma} while analysing a similar new physics case.  }

The proposed scenario also gives some predictions in the scalar sector relevant for collider searches.    
Neglecting mixing between the scalar bidoublet $\Phi$ and the Higgs doublets $\phi^{(\prime)}$, the scalar spectrum will contain a heavy CP-even neutral scalar transforming as an SU(2)$_L$ singlet originating from $\Phi$.  We will denote this state by $h_2$. The mass of this scalar is expected to be around the symmetry breaking scale $u \sim$~TeV.  The dominant interactions of $h_2$ are with the heavy fermions and the heavy gauge vector bosons, these are described by
\begin{align} \label{hghg}
\mathcal{L} \supset& \;  2 (     M_{W^{\prime}}^2 W_{\mu}^{\prime +}   W^{\prime - \mu}   + \frac{1}{2}   M_{Z^{\prime}}^2   Z_{\mu}^{\prime}  Z^{\prime \mu} )   \frac{h_2}{u}  - (y_{Q})_{ii}  \, \bar Q_i Q_i  \, h_2  - (y_{L})_{ii}  \, \bar L_i L_i  \, h_2   \,,
\end{align}
with $Q_i^T = (U_i, D_i)$, $L_i^T = (N_i, E_i)$  ($i=1,2$) and
\begin{align}
y_{Q} =  \frac{g_2^2}{n_1^2}
\begin{pmatrix} 
  \widetilde M_{Q_1}  &   0  \\
0 &  \widetilde M_{Q_2}  (  \Delta_{s}^2 + \Delta_b^2   )
\end{pmatrix} \,, \qquad
y_{L} =  \frac{g_2^2}{n_1^2}
\begin{pmatrix} 
  \widetilde M_{L_1}  &   0  \\
0 &  \widetilde M_{L_2}  (  \Delta_{\mu}^2 + \Delta_{\tau}^2   ) 
\end{pmatrix}  \,.
\end{align}
The production of $h_2$ at the LHC is dominated by gluon fusion mediated by the heavy quarks and is determined by the same parameters entering in the low-energy global fit.    At the centre-of mass energy $\sqrt{s}$ the production cross-section reads
\begin{align}
\sigma(pp \to h_2) \simeq  \frac{c_{gg}  \Gamma(h_2 \to gg)}{M_{h_2} s} \,, \qquad \Gamma(h_2 \to gg) \simeq  \frac{\alpha_s^2 M_{h_2}^3}{18 \pi^3}  \left|  \sum_{i=1}^2 \frac{(y_{Q})_{ii}}{u  \widetilde M_{Q_i} }  \right|^2  \,.
\end{align}
Here $c_{gg}$ represents a dimensionless partonic integral which we estimate using the set of parton distribution functions MSTW2008NLO~\cite{Martin:2009iq} evaluated at the scale $\mu = M_{h_2}$.   In writing the decay rate for $h_2 \to gg$ we have taken the local approximation for the fermionic loops.    For $M_{h_2} \sim 1$~TeV, and restricting the rest of the parameters to the region described in Eq.~\eqref{exkfn}, we obtain $\sigma(pp \to h_2) \simeq 110-290$~fb at $\sqrt{s} =13$~TeV centre-of-mass energy. For $M_{Z^{\prime}} \sim 1.7$~TeV (and $M_{h_2} \sim 1$~TeV) the production cross-section converges towards $\sim 110$~fb.   The interactions of $h_2$ in \Eq{hghg} will induce loop-mediated decays into gluons (which will hadronize into jets) and electroweak gauge bosons $W^+ W^-$, $ZZ$, $\gamma \gamma$, $Z \gamma$.  Assuming negligible tree-level decays, the $h_2$ boson will manifest in this case as a very narrow resonance decaying mainly into a pair of jets. The current experimental sensitivity for dijet-resonances at the LHC around this mass range ($M_{h_2} \sim 1$~TeV) is at the level of $10^3$~fb~\cite{ATLAS:2015nsi,Khachatryan:2015dcf}.    The decays into electroweak gauge bosons are found to be subdominant and for $M_{Z^{\prime}} \in [1, 1.7]$~TeV we have: $\mathrm{Br}(h_2 \to WW) \sim  10^{-2}$, $\mathrm{Br}(h_2 \to ZZ, Z \gamma)/\mathrm{Br}(h_2 \to WW) \sim 25\%$, $\mathrm{Br}(h_2 \to \gamma \gamma)/\mathrm{Br}(h_2 \to WW) \sim 1\%$.     Note however that in the case where some of the heavy fermions are below the threshold $M_{h_2}/2$, tree-level decay of $h_2$ into these fermions becomes kinematically open and will generically dominate over the loop-induced decays commented above.

\section{Conclusions}
\label{sec:conclusions}

We have performed a phenomenological analysis of a renormalizable and perturbative gauge extension of the Standard Model.
We took into account flavour observables sensitive to tree-level new physics contributions as well as bounds from electroweak precision measurements at the $Z$ and $W$ pole.  More specifically, we have analysed the model in light of the current hints of new physics in $b \to c \ell \nu$ and $b \to s \ell^+ \ell^-$ semileptonic decays, finding that the flavour anomalies can  be accommodated within the allowed regions of the parameter space.

As derived from the phenomenological analysis, strong hierarchies in the flavour structure of the Yukawa couplings are required in order to accommodate both $b \to s \ell^+ \ell^-$ and $b \to c \ell \nu$ anomalies.   We have taken a phenomenologically oriented approach in this work,  not invoking any flavour symmetry behind such structure.   One interesting question would be the exploration of possible flavour symmetries accommodating the observed flavour structure.    We confirm the conclusions of Ref.~\cite{Boucenna:2016wpr} regarding the importance of suppressing gauge bosons mixing. This translates in a tuning of $\tan \beta$.  Such accidental tuning would be more satisfactory if there was a dynamical or symmetry-based explanation behind.    These last points also bring us to the question of the validity of our analysis, based on tree-level new physics effects, once quantum corrections are considered.  These corrections might alter the flavour structure of the theory, remove accidental tunings which hold at the classical level as well as introduce new constraints from loop-induced processes such as $b \to s \gamma$.  Though such analysis lies beyond the scope of our work, it would be relevant in order to establish the viability of the proposed framework if the present deviations in $b \to s \ell^+ \ell^-$ and $b \to c \ell \nu$ are confirmed in the future.

From the model building point of view, there are many open questions which we have not addressed in this work and would deserve further investigation, one of them being the implementation of a mechanism for the generation of the observed neutrino masses and lepton mixing angles.   Our model also lacks a dark matter candidate, motivating the extension of our framework.  It would be interesting to pursue the investigation of possible embeddings of the model within a larger gauge group, where the mass of the heavy fermions arise from spontaneous symmetry breaking.

\subsubsection*{Acknowledgements}
\linespread{1}\selectfont
{\small
We thank S\'ebastien Descotes-Genon, Thorsten Feldmann, Martin Jung, Admir
Greljo, and the participants of the June 2016 BaBar Collaboration Meeting at SLAC
for helpful discussions.  We also thank Adam Falkowski for providing
numerical results used in our analysis of electroweak precision data.
S.M.B. acknowledges support of the MIUR grant for the Research
Projects of National Interest PRIN 2012 No.2012CP-PYP7 Astroparticle
Physics, of INFN I.S. TASP2014, and of MultiDark CSD2009-00064.  The
work of A.C. is supported by the Alexander von Humboldt Foundation.
The work of J.F. is supported in part by the Spanish Government, ERDF
from the EU Commission and the Spanish Centro de Excelencia Severo
Ochoa Programme [Grants~No.~FPA2011-23778, FPA2014-53631-C2-1-P,
  PROMETEOII/2013/007 and SEV-2014-0398].  J.F.  is also supported by 
  an ``Atracci\'{o} de Talent" scholarship from VLC-CAMPUS.
A.V. acknowledges financial support from the ``Juan de la Cierva''
program (27-13-463B-731) funded by the Spanish MINECO as well as from
the Spanish grants FPA2014-58183-P, Multidark CSD2009-00064,
SEV-2014-0398 and PROMETEOII/ 2014/084 (Generalitat Valenciana).
J.V. is funded by the Swiss National Science Foundation. J.V. acknowledges support from Explora project FPA2014-61478-EXP.
We have used {\tt SARAH}~\cite{Staub:2013tta} and {\tt SPheno}~\cite{Porod:2003um,Porod:2011nf} 
to cross-check some of our results, and {\tt matplotlib} to produce most of the 
plots in the paper~\cite{Hunter:2007}.   
}

\newpage

\appendix

\section{Details of the Model}
\label{sec:modeldetails}

\subsection{Tadpole equations}
\label{subsec:tadpole}

The vev configuration introduced in Section \ref{subsec:scalar} leads to
three minimization conditions or tadpole equations. In the following we
will consider all the parameters in the scalar potential to be
real. Defining
\begin{equation}
t_i = \frac{\partial \mathcal V}{\partial {v}_i} = 0 \, ,
\end{equation}
these are
\begin{align}   \label{eq:tadpole}
\begin{aligned}
t_\phi &= m_\phi^2 v_\phi + \frac{1}{2} v_\phi \left( \lambda_4 v_{\phi^\prime}^2 + \lambda_5 u^2 \right) + \frac{1}{2} v_{\phi^\prime} u \, \mu + \frac{1}{2} \lambda_1 v_\phi^3 \, , \\ 
t_{\phi^\prime} &=  m_{\phi^\prime}^2 v_{\phi^\prime} + \frac{1}{2} v_{\phi^\prime} \left( \lambda_4 v_\phi^2 + \lambda_6 u^2 \right) + \frac{1}{2} v_\phi u \, \mu + \frac{1}{2} \lambda_2 v_{\phi^\prime}^3\, , \\ 
t_u &= m_{\Phi}^2 u + \frac{1}{2} u \left( \lambda_5 v_\phi^2 + \lambda_6 v_{\phi^\prime}^2 \right) + \frac{1}{2} v_\phi v_{\phi^\prime} \mu + \frac{1}{2} \lambda_3 u^3 \, .  \end{aligned}
\end{align}
These three conditions can be solved for the mass squared parameters
$m_\phi^2$, $m_{\phi^\prime}^2$ and $m_\Phi^2$.

\subsection{Scalar mass matrices}
\label{subsec:mass}

The neutral scalar fields can be decomposed as
\begin{align}
\begin{aligned}
\varphi^0 &= \frac{1}{\sqrt{2}} \left( v_\phi + S_\phi + i \, A_\phi \right) \, , \\
\varphi'^0 &= \frac{1}{\sqrt{2}} \left( v_{\phi^\prime} + S_{\phi^\prime} + i \, A_{\phi^\prime} \right) \, , \\
\Phi^0 &= \frac{1}{\sqrt{2}} \left( u + S_{\Phi} + i \, A_{\Phi} \right) \, .
\end{aligned}
\end{align}
Since we assume that CP is conserved in the scalar sector, the CP-even and CP-odd
states do not mix. In this case, one can define the bases
\begin{align}
\begin{aligned}
 \Ss^T & \equiv \left( S_\phi, S_{\phi^\prime}, S_\Phi \right) \quad , \qquad &\Pp^T& \equiv \left( A_\phi, A_{\phi^\prime}, A_\Phi \right) \, , \\[3mm]
 (\Hh^{-})^T & \equiv \left( \left(\varphi^+\right)^\ast , \left(\varphi'^+\right)^\ast , \left(\Phi^+\right)^\ast \right) \quad , \qquad &(\Hh^{+})^T& \equiv \left( \varphi^+ , \varphi'^+ , \Phi^+ \right) \, ,
\end{aligned}
\end{align}
which allow us to obtain the scalar mass Lagrangian
\begin{equation}
-\L_{m}^s = \frac{1}{2} \Ss^T \M_{\Ss}^2 \Ss + \frac{1}{2} \Pp^T \M_{\Pp}^2 \Pp + \left(\Hh^-\right)^T \M_{\Hh^\pm}^2 \Hh^+ \, .
\end{equation}
The mass matrix for the CP-even scalars is given by
\begin{equation} 
\M_{\Ss}^2 = \left( 
\begin{array}{ccc}
\M_{S_\phi S_\phi}^2 &\M_{S_\phi S_{\phi^\prime}}^2 & \M_{S_\phi S_\Phi}^2 \\ 
\M_{S_\phi S_{\phi^\prime}}^2 & \M_{S_{\phi^\prime} S_{\phi^\prime}}^2 & \M_{S_{\phi^\prime} S_\Phi}^2 \\ 
\M_{S_\phi S_\Phi}^2 & \M_{S_{\phi^\prime} S_\Phi}^2 & \M_{S_\Phi S_\Phi}^2 
\end{array} 
\right) \, ,
\end{equation} 
with
\begin{align} 
\begin{aligned}
\M_{S_\phi S_\phi}^2 &= m_\phi^2 + \frac{1}{2} \Big(3 v_{\phi}^{2} {\lambda}_{1} + v_{\phi^\prime}^2 {\lambda}_{4} + u^2 {\lambda}_{5}  \Big) \, , \\ 
\M_{S_\phi S_{\phi^\prime}}^2 &= v_{\phi} v_{\phi^\prime} {\lambda}_{4} + \frac{1}{2} u \mu \, , \\ 
\M_{S_\phi S_\Phi}^2 &= v_{\phi} u \lambda_5 + \frac{1}{2} v_{\phi^\prime} \mu \, , \\ 
\M_{S_{\phi^\prime} S_{\phi^\prime}}^2 &= m_{\phi^\prime}^2 + \frac{1}{2} \Big(3 v_{\phi^\prime}^{2} {\lambda}_{2} + v_{\phi}^2 {\lambda}_{4} + u^2 {\lambda}_{6}  \Big) \, , \\  
\M_{S_{\phi^\prime} S_\Phi}^2 &= v_{\phi^\prime} u \lambda_6  + \frac{1}{2} v_{\phi} \mu \, , \\ 
\M_{S_\Phi S_\Phi}^2 &= {m}_{\Phi}^{2} + \frac{1}{2} \Big( v_{\phi}^{2} {\lambda}_{5}  + v_{\phi^\prime}^2 {\lambda}_{6} + 3 u^2 \lambda_3 \Big) \, .
\end{aligned}
\end{align} 
The lightest CP-even state, $\Ss_1 \equiv h$, is identified with
the recently discovered SM-like Higgs boson with a mass $\sim 125$
GeV. Similarly, in the Landau gauge ($\xi = 0$), the mass matrix for
the CP-odd scalars is given by
\begin{equation} 
\M_{\Pp}^2 = \left( 
\begin{array}{ccc}
\M_{A_\phi A_\phi}^2 &\M_{A_\phi A_{\phi^\prime}}^2 & \M_{A_\phi A_\Phi}^2 \\ 
\M_{A_\phi A_{\phi^\prime}}^2 & \M_{A_{\phi^\prime} A_{\phi^\prime}}^2 & \M_{A_{\phi^\prime} A_\Phi}^2 \\ 
\M_{A_\phi A_\Phi}^2 & \M_{A_{\phi^\prime} A_\Phi}^2 & \M_{A_\Phi A_\Phi}^2
\end{array} 
\right) \, ,
\end{equation} 
with
\begin{align} 
\begin{aligned}
\M_{A_\phi A_\phi}^2 &= m_\phi^2 + \frac{1}{2} \Big(v_{\phi}^{2} {\lambda}_{1} + v_{\phi^\prime}^2 {\lambda}_{4} + u^2 {\lambda}_{5} \Big) \, , \\ 
\M_{A_\phi A_{\phi^\prime}}^2 &= \frac{1}{2} u \mu \, , \\ 
\M_{A_\phi A_\Phi}^2 &=  \frac{1}{2} v_{\phi^\prime} \mu \, , \\
\M_{A_{\phi^\prime} A_{\phi^\prime}}^2 &= m_{\phi^\prime}^2 + \frac{1}{2} \Big(v_{\phi^\prime}^{2} {\lambda}_{2} + v_{\phi}^2 {\lambda}_{4} + u^2 {\lambda}_{6} \Big) \, , \\  
\M_{A_{\phi^\prime} A_\Phi}^2 &= -  \frac{1}{2} v_{\phi} \mu \, , \\ 
\M_{A_\Phi A_\Phi}^2 &= {m}_{\Phi}^{2} + \frac{1}{2} \Big( v_{\phi}^{2} {\lambda}_{5}  + v_{\phi^\prime}^2 {\lambda}_{6} + u^2 \lambda_3 \Big) \, .
\end{aligned}
\end{align} 
After application of the tadpole equations in
Eq.~\eqref{eq:tadpole}, it is straightforward
to show that the matrix $\M_{\Pp}^2$ has two vanishing
eigenvalues. These correspond to the Goldstone bosons that constitute
the longitudinal modes for the massive $Z$ and $Z^\prime$
bosons. Finally, the mass matrix for the charged scalars in the Landau
gauge ($\xi = 0$) is given by
\begin{equation} 
\M_{\Hh^\pm}^2 = \left( 
\begin{array}{ccc}
\M_{\varphi^+ \varphi^+}^2 &\M_{\varphi^+ \varphi'^+}^2 & \M_{\varphi^+ \Phi^+}^2 \\ 
\M_{\varphi^+ \varphi'^+}^2 & \M_{\varphi'^+ \varphi'^+}^2 & \M_{\varphi'^+ \Phi^+}^2 \\ 
\M_{\varphi^+ \Phi^+}^2 & \M_{\varphi'^+ \Phi^+}^2 & \M_{\Phi^+ \Phi^+}^2
\end{array} 
\right) \, ,
\end{equation} 
with
\begin{align} 
\begin{aligned}
\M_{\varphi^+ \varphi^+}^2 &= m_\phi^2 + \frac{1}{2} \Big(v_{\phi}^{2} {\lambda}_{1} + v_{\phi^\prime}^2 {\lambda}_{4} + u^2 {\lambda}_{5} \Big) \, , \\ 
\M_{\varphi^+ \varphi'^+}^2 &= \frac{1}{2} u \mu \, , \\ 
\M_{\varphi^+ \Phi^+}^2 &=    - \frac{1}{2} v_{\phi^\prime} \mu \, , \\ 
\M_{\varphi'^+ \varphi'^+}^2 &= m_{\phi^\prime}^2 + \frac{1}{2} \Big(v_{\phi^\prime}^{2} {\lambda}_{2} + v_{\phi}^2 {\lambda}_{4} + u^2 {\lambda}_{6} \Big) \, , \\  
\M_{\varphi'^+ \Phi^+}^2 &= \frac{1}{2} v_{\phi} \mu \, , \\
\M_{\Phi^+ \Phi^+}^2 &= {m}_{\Phi}^{2} + \frac{1}{2} \Big( v_{\phi}^{2} {\lambda}_{5}  + v_{\phi^\prime}^2 {\lambda}_{6} + u^2 \lambda_3 \Big) \, .
\end{aligned}
\end{align} 
Again, one can find two vanishing eigenvalues in $\M_{\Hh^\pm}^2$ after
applying the tadpole equations in
Eqs. \eqref{eq:tadpole}. These correspond to
the Goldstone bosons \emph{eaten-up} by the $W$ and $W^\prime$ gauge
bosons.

\section{Pseudo-observables for $Z$- and $W$-pole observables}
\label{sec:EWPD}

In our model, the pseudo-observables considered in Ref.~\cite{Efrati:2015eaa} are given by:
\begin{align}
\begin{aligned}
\delta m &= - \delta v\, \frac{g^{\prime 2}}{g^2 - g^{\prime 2}} \ ,\\[2pt]
\delta g_L^{W\ell_i}&= - \zeta\, \ep2\, \frac{g_2^4}{n_1^4}  \Delta^\ell_{ii}+f(1/2,0)-f(-1/2,-1)\,,\\[2pt]
\delta g_L^{Z\ell_i}&= \zeta\, \ep2\,\frac{g_2^4}{2n_1^4}  \Delta^\ell_{ii}+f(-1/2,-1)\,,\\[2pt]
 \delta g_R^{Z\ell_i}&=f(0,-1)\,,\\[2pt]
\delta g_L^{Zu_i}&=-\zeta\, \ep2 \frac{g_2^4}{2n_1^4} (V_{\rm{CKM}}\Delta^q V_{\rm{CKM}}^\dagger)_{ii}+f(1/2,2/3)\,,\\[2pt]
 \delta g_R^{Zu_i}&=f(0,2/3)\,,\\[2pt]
\delta g_L^{Zd_i}&=\zeta\, \ep2 \frac{g_2^4}{2n_1^4}  \Delta^q_{ii}+f(-1/2,-1/3)\,,\\[2pt]
\delta g_R^{Zd_i}&= f(0,-1/3)\,,
\end{aligned}
\end{align}
where
\eq{
\delta v=- \zeta \ep2\frac{1}{2}\frac{g_2^4}{n_1^4} \Delta^\ell_{22}\quad \text{and}\quad
f(T^3,Q)=-\delta v\left(T^3+Q\frac{g'^{\,2}}{g^2-g'^{\,2}}\right)\,.
}
The family index $i$ for these shifts covers the three fermion
generations except for $\delta g_R^{Zu_i}$, for which $i=1,2$. We
neglect corrections to the right-handed $Z$ and $W$ couplings that are
suppressed by the fermion masses, see~\Sec{sec:strategy}. We also
neglect loop contributions, which we estimate to be comparable to the
tree-level contributions for $\zeta \lesssim 0.02$. However, the
resulting $\delta g$'s in that case would be below the limits
quoted in \cite{Efrati:2015eaa}.

\bibliographystyle{JHEP}
\bibliography{refs}

\end{document}